\documentclass[fleqn,usenatbib]{mnras}

\usepackage{newtxtext,newtxmath}

\usepackage[T1]{fontenc}

\DeclareRobustCommand{\VAN}[3]{#2}
\let\VANthebibliography\thebibliography
\def\thebibliography{\DeclareRobustCommand{\VAN}[3]{##3}\VANthebibliography}

\usepackage{graphicx}	
\usepackage{amsmath}	
\usepackage{xcolor}
\usepackage{mwe}
\usepackage{subfig}
\usepackage[normalem]{ulem}
\usepackage{tabularx}
\usepackage[roman]{parnotes}
\usepackage{mathtools}
\usepackage{multirow}
\usepackage[thinc]{esdiff}
\usepackage{physics}
\usepackage{cleveref}

\newcommand{\aperp}{\alpha_\perp}	
\newcommand{\apar}{\alpha_\parallel}

\title[Geometry and Growth from Voids]{Measurements of cosmic expansion and growth rate of structure from voids in the Sloan Digital Sky Survey between redshift 0.07 and 1.0}

\author[A. Woodfinden et al.]{Alex Woodfinden$^{1,2}$\thanks{E-mail: awoodfin@uwaterloo.ca}, Seshadri Nadathur$^{3}$, Will J. Percival$^{1,2,4}$, Slađana Radinović$^{5}$, Elena Massara$^{1,2}$, \newauthor{Hans A. Winther$^{5}$}
\\
$^{1}$Waterloo Centre for Astrophysics, University of Waterloo, 200 University Ave W, Waterloo, ON N2L 3G1, Canada\\
$^{2}$Department of Physics and Astronomy, University of Waterloo, 200 University Ave W, Waterloo, ON N2L 3G1, Canada\\
$^{3}$Institute of Cosmology and Gravitation, University of Portsmouth, Burnaby Road, Portsmouth, PO1 3FX, United Kingdom\\
$^{4}$Perimeter Institute for Theoretical Physics, 31 Caroline St North, Waterloo, ON N2L 2Y5, Canada\\
$^{5}$Institute of Theoretical Astrophysics, University of Oslo, P.O. Box 1029 Blindern, N-0315 Oslo, Norway
}

\date{Accepted XXX. Received YYY; in original form ZZZ}

\pubyear{2021}

\begin{document}
\label{firstpage}
\pagerange{\pageref{firstpage}--\pageref{lastpage}}
\maketitle

\begin{abstract}
We present measurements of the anisotropic cross-correlation of galaxies and cosmic voids in data from the Sloan Digital Sky Survey Main Galaxy Sample (MGS), Baryon Oscillation Spectroscopic Survey (BOSS) and extended BOSS (eBOSS) luminous red galaxy catalogues from SDSS Data Releases 7, 12 and 16, covering the redshift range $0.07<z<1.0$. As in our previous work analysing voids in subsets of these data, we use a reconstruction method applied to the galaxy data before void-finding in order to remove selection biases when constructing the void samples. We report results of a joint fit to the multipole moments of the measured cross-correlation for the growth rate of structure, $f\sigma_8(z)$, and the ratio $D_\mathrm{M}(z)/D_\mathrm{H}(z)$ of the comoving angular diameter distance to the Hubble distance, in six redshift bins. For $D_\mathrm{M}/D_\mathrm{H}$, we are able to achieve a significantly higher precision than that obtained from analyses of the baryon acoustic oscillations (BAO) and galaxy clustering in the same datasets. Our growth rate measurements are of lower precision but still comparable with galaxy clustering results. For both quantities, the results agree well with the expectations for a $\Lambda$CDM model. Assuming a flat Universe, our results correspond to a measurement of the matter density parameter $\Omega_\mathrm{m}=0.337^{+0.026}_{-0.029}$. For more general models the degeneracy directions obtained are consistent with and complementary to those from other cosmological probes. These results consolidate void-galaxy cross-correlation measurements as a pillar of modern observational cosmology.
\end{abstract}




\section{Introduction}

The best current evidence for the standard $\Lambda$ Cold Dark Matter ($\Lambda$CDM) cosmological model relies on the combination of Cosmic Microwave Background observations by the Planck satellite \citep{Planck2020} together with observations at lower redshift. The most robust low-redshift measurements come from the Baryon Acoustic Oscillations (BAO), which use the relics of primordial sound waves seen in the distribution of galaxies as a standard ruler \citep{Alam-BOSS:2017,Alam-eBOSS:2021}. Future galaxy surveys including DESI \citep{desi1,desi2} and Euclid \citep{Euclid} that are designed to observe the BAO feature require redshifts for large numbers of galaxies over large volumes. These surveys also allow other cosmological measurements, including those from redshift-space distortions in the galaxy field (RSD; \citealt{Kaiser:1987}), and from the distribution of galaxies around voids \citep{Lavaux:2012}. The latter is the focus of our study. 

Voids are interesting objects to study because the link between the nonlinear density into the nonlinear velocity can be accurately mapped using linear theory \citep{Paz:2013,Cai:2016a,Nadathur:2019a}. As a consequence, the RSD signal in the distribution of galaxies around voids can be analytically modelled to small scales, and we can obtain information to smaller scales from the RSD and AP measurements, than if we had tried to model all galaxy pairs \citep{Lavaux:2012,Hamaus:2016,Nadathur:2019c}. There have consequently been many studies of the AP and RSD effects using the void-galaxy correlation \citep{Paz:2013,Hamaus:2016,Hamaus:2017a,Hawken:2017,Nadathur:2019c,Achitouv:2019,Hawken:2020,Aubert20a} and closely related statistics \citep{Paillas:2021}. In principle, the statistical precision with which $D_\mathrm{M}(z)/D_\mathrm{H}(z)$ can been measured using voids exceeds that obtained from BAO \citep{Hamaus:2016,Nadathur:2019c}, although the potential for systematics is slightly higher given the need to model the RSD signal. In addition there are other ways in which voids can be used to test cosmological models including using the void size distribution, void lensing, or void-void clustering \citep[e.g.][]{Pisani:2015,Sanchez:2016,Nadathur:2016a,Raghunathan:2020,Zhao:2021}.

Key to the geometrical constraints provided at low redshift by voids is the dilation of clustering caused by the distance-redshift relationship applied to convert redshifts into comoving distances. Along the line-of-sight, provided that the clustering is measured on scales that are small compared to those over which cosmological evolution occurs, the clustering is sensitive to $D_\mathrm{H}(z)\equiv c/H(z)$, where $H(z)$ is the redshift-dependent Hubble parameter. Across the line-of-sight we are sensitive to the comoving angular diameter distance $D_\mathrm{M}(z)$, where \citep{Hogg:1999}\footnote{Care needs to be taken when evaluating this expression numerically at $\Omega_\mathrm{K}=0$ and $\Omega_\mathrm{K} < 0$. For $\Omega_\mathrm{K}=0$ one finds $\lim_{\Omega_\mathrm{K}\to0} D_\mathrm{M} = D_\mathrm{C}$ and for $\Omega_\mathrm{K}<0$ one finds $D_\mathrm{H}(0)\frac{1}{\sqrt{\Omega_\mathrm{K}}}\sinh\left(\sqrt{\Omega_\mathrm{K}} \frac{D_\mathrm{C}}{D_\mathrm{H}(0)}\right) = D_\mathrm{H}(0)\frac{1}{\sqrt{- \Omega_\mathrm{K}}}\sin\left(\sqrt{- \Omega_\mathrm{K}} \frac{D_\mathrm{C}}{D_\mathrm{H}(0)}\right).$}
\begin{equation}
    \label{eq:DM}
    D_\mathrm{M} = D_\mathrm{H}(0)\frac{1}{\sqrt{\Omega_\mathrm{K}}}\sinh\left(\sqrt{\Omega_\mathrm{K}} \frac{D_\mathrm{C}}{D_\mathrm{H}(0)}\right),
\end{equation}
and the (line-of-sight) comoving distance is
\begin{equation}
    \label{eq:DC}
    D_\mathrm{C}(z) \equiv\int_0^z dz' \frac{c}{H(z')}\,.
\end{equation}
Knowing that the clustering is isotropic, we will only recover this in our comoving maps if we use the true value of $D_\mathrm{M}(z)/D_\mathrm{H}(z)$ when converting redshifts to distances (in the absence of other effects). Separate measurements of $D_\mathrm{H}(z)$ and $D_\mathrm{M}(z)$ can be made if we have a standard ruler whose intrinsic length we know or that depends on other cosmological parameters, such as the BAO scale. Whereas to measure the dimensionless ratio $D_\mathrm{M}(z)/D_\mathrm{H}(z)$ we only need an object - such as a stack of voids - that we can use as a standard shape, knowing that it is expected to be spherical but not knowing its intrinsic size (called the AP effect, \citealt{Alcock:1979}). 

In general, these geometrical measurements are degenerate with RSD which also cause anisotropic distortions in the derived maps. This is, however, not true for the BAO position as the BAO in redshift space are at the same locations as in real-space. For voids, we can distinguish RSD and AP because they affect the apparent shape in different ways \citep{Nadathur:2019c}. The measurements of the AP and RSD effects from voids are not strongly correlated with those obtained from analyses of galaxy-galaxy clustering \citep{Nadathur:2020a}, so they represent additional information that can be obtained from existing surveys. 

In this paper we build on previous work developed for the cosmological analysis of voids to analyse galaxy samples within the Sloan Digital Sky Survey (SDSS-II; \citealt{York:2000}). We analyse the Main Galaxy Sample (MGS; \citealt{Howlett:2015,Ross:2015}), the Baryon Oscillation Spectroscopic Survey \citep[BOSS;][]{Dawson:2013} of SDSS-III \citep{Eisenstein:2011}, and the extended BOSS \citep[eBOSS;][]{Dawson:2016} of SDSS-IV \citep{Blanton:2017}, covering a wide range in redshift using a single analysis method for the first time. These data represent the best public galaxy redshift survey data available to date, and will only be surpassed when DESI \citep{desi1,desi2} \& Euclid \citep{Euclid} results are released. The analysis method used is consistent with that applied to eBOSS in \citet{Nadathur:2020b}, and is a development of that used for BOSS by \citet{Nadathur:2019a}. It has not previously been applied to the MGS or low redshift BOSS samples. By analysing these new data and consolidating previous analyses, we are able to build a picture of the geometrical evolution of the Universe and the evolution of the growth of structure within it from only the analysis of SDSS galaxies around voids. In this work we do not consider the additional eBOSS samples of quasars and emission line galaxies (ELGs) that extend out to higher redshifts ($z < 2.2$). The sparsity of tracers in the quasar sample means that reconstruction technique our method relies on is not efficient. On the other hand, the ELG sample was selected from imaging data that had anisotropic properties and suffers from significant angular fluctuations \citep{Raichoor:2021}. \citet{Tamone2020,deMattia:2021} showed that careful corrections for these effects could be made for BAO and RSD analyses of galaxy clustering, but we leave extensions of this work to the void-galaxy correlation to future work.

Our paper is structured as follows: we introduce the data and mock catalogues analysed in Section~\ref{sec:data}. In addition to the MGS and BOSS data, we also make use of mock galaxy catalogues to test our analysis pipeline and to estimate the statistical errors for our data measurements (described in sections~\ref{sec:Patchy Mocks} \&~\ref{sec:MGS Mocks}). A smaller collection of full $N$-body mocks are used, in addition to the approximate mocks, to quantify the magnitude of the systematic errors (described in section~\ref{sec:NSERIES Mocks}). Finally we use mocks created from full $N$-body simulation boxes with dark matter information to calibrate template profiles used in the theoretical modelling (described in Section~\ref{sec:BigMD Mocks}). We review the analysis pipeline in Section~\ref{sec:analysis}. Section~\ref{sec:systematic} outlines a number of tests performed to confirm that the analysis pipeline is robust and accurate. The results of our analyses are presented in Section~\ref{sec:result} and are discussed in Section~\ref{sec:conclusions}.

\section{Data and Mocks} 
\label{sec:data}

\subsection{MGS}
\label{sec:MGS}

The Main Galaxy Sample \citep[MGS;][]{Strauss2002} is a selection of galaxies from the SDSS-I and SDSS-II surveys \citep{York:2000} Data Release 7 \citep[DR7; ][]{Abazajian:2009}, at redshifts $z<0.2$, with spectra taken using spectrographs mounted on the 2.5-meter telescope at Apache Point Observatory in New Mexico \citep{Gunn2006}. A subsample of this catalogue, created for large-scale structure analyses, is described by \citet{Ross:2015} and \citet{Howlett:2015}, 
which used additional colour, magnitude and redshift cuts to obtain a high-bias ($b\sim1.5$) sample of galaxies occupying dark matter halos with $M_\mathrm{halo}>10^{12}\,M_\odot$, and with a high galaxy density. This sample, which we refer to as MGS, consists of 62\,163 galaxies covering a contiguous footprint of 6813\,deg$^2$ in the Northern Galactic Cap (NGC) region between redshifts $0.07<z<0.2$. The MGS footprint is shown in Figure~\ref{fig:footprints}. Systematic weights are included in the catalogue to correct for angular fluctuations due to target selection effects \citep{Ross:2012}.

\subsubsection{MGS Mocks}
\label{sec:MGS Mocks}

We use 1000 mock galaxy catalogues matching the footprint, redshift distribution and clustering properties of the MGS data \citep{Howlett:2015}. These mocks were built from 500 independent dark matter simulations at $z=0.15$ created using the \texttt{PICOLA} algorithm \citep{Howlett2015b}, with fiducial cosmology $\Omega_\mathrm{m}=0.31$, $\Omega_b=0.048$, $h = 0.67$, $\sigma_8 = 0.83$ and $n_s = 0.96$. Halos were selected in this field using a friends-of-friends algorithm, and populated with mock galaxies using a Halo Occupation Distribution (HOD) prescription with parameters fitted to the MGS data, as described in \citet{Ross:2015}. From each box two non-overlapping sections were then cut out to match the MGS footprint, and the mocks subsampled to match the redshift-dependence of the mean galaxy density in the data. We use all 1000 of these mocks to obtain accurate estimate of the covariance matrix for the measurement, and use a subset of 250 of them to test our analysis methods for systematic errors. 

\subsection{BOSS}
\label{sec:boss}

The Baryon Oscillation Spectroscopic Survey \citep[BOSS][]{Dawson:2013} of SDSS-III \citep{Eisenstein:2011} measured spectra from more than 1.5 million objects using spectrographs upgraded from those used for MGS, mounted on the 2.5-meter Sloan telescope \citep{Gunn2006}. The target sample covered nearly 10\,000 deg$^2$ of the sky over two hemispheres, the North Galactic Cap (NGC) and the South Galactic Cap (SGC). 

\begin{figure*}
    \centering
    \includegraphics[width=0.85\textwidth]{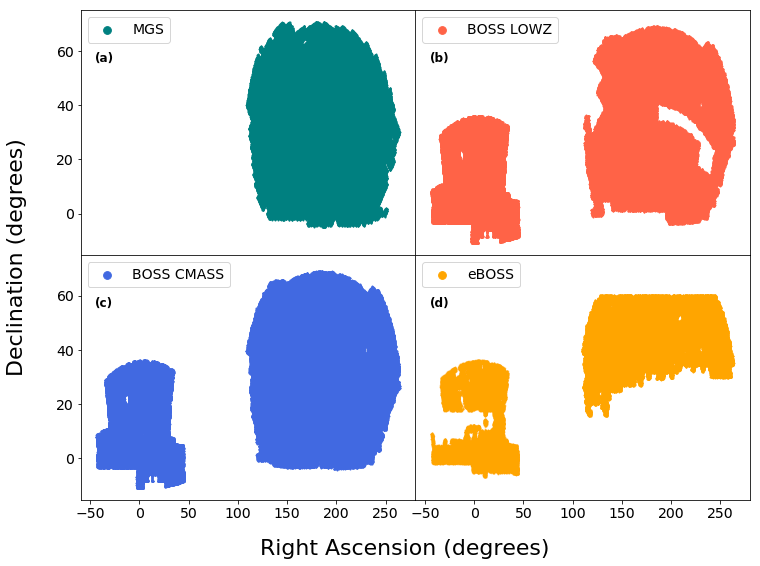}
    \caption{Footprints of the different SDSS galaxy catalogues: (a) the MGS from SDSS DR7, covering $0.07<z<0.2$; (b) the BOSS LOWZ catalogue from SDSS DR12, covering $0.2<z\lesssim0.43$; (c) the BOSS CMASS catalogue from SDSS DR12, covering $0.4\lesssim z\lesssim0.75$; (d) the eBOSS LRG catalogue from SDSS DR16, covering $0.6<z<1.0$. We treat the MGS separately from the others, but the different BOSS and eBOSS samples overlap with each other and are combined in our analysis in order to use the data more efficiently. As a result the changes in the survey mask with redshift must be accommodated in the void-finding, as described in Section~\ref{sec:voidfinding}.}
    \label{fig:footprints} 
\end{figure*}

The final BOSS data were included in Data Release 12 \citep[DR12][]{Alam-DR11&12:2015}. The large-scale structure catalogues were created using two different target selection algorithms, LOWZ and CMASS \citep{Reid2016}. The LOWZ sample was designed to target luminous red galaxies (LRGs) in the redshift range $0.2 \lesssim z \lesssim 0.4$, while the CMASS sample was designed to target LRGs in a narrow mass range at redshifts $0.4 \lesssim z \lesssim 0.75$. The LOWZ footprint is slightly smaller than, and fully encompassed within, the CMASS footprint, as shown in Figure~\ref{fig:footprints}. Despite these small differences, the LOWZ and CMASS samples show very similar clustering amplitudes across both NGC and SGC, and we follow \citet{Alam-BOSS:2017} in analysing the \emph{combined sample}, including a small region of redshift overlap. As described below in Section~\ref{sec:voidfinding}, the change in the survey footprint around $z\simeq0.43$ must be accounted for when identifying voids, but allows for a more efficient use of the data.

In the recent eBOSS Data Release 16 \citep[DR16;][]{Ahumada:2020} cosmological analyses \citep{Alam-eBOSS:2021}, the high-redshift end of the CMASS sample at $z>0.6$ is combined with the eBOSS LRG sample, which overlaps with it in redshift and footprint. A measurement of the void-galaxy cross-correlation in this combined eBOSS+CMASS sample has already been presented by \citet{Nadathur:2020b}, using very similar methods to those we use here. We therefore restrict our analysis in this work to the section of the combined BOSS LOWZ and CMASS samples that are below this redshift. In practice, to minimize the loss of voids due to edge effects close to a survey boundary, we use all galaxies with $0.2<z<0.63$ for void-finding, but then select only those voids whose centres lie at $z<0.6$ for the cross-correlation measurement.  

The void-galaxy cross-correlation in a subset of these data, corresponding to the CMASS sample alone in the range $0.43<z<0.7$, was analysed by \citet{Nadathur:2019c}. That work used a single redshift bin and reported very precise constraints on $f\sigma_8$ and $D_\mathrm{M}/D_\mathrm{H}$ at the single effective redshift $z_\mathrm{eff}=0.57$. However, the CMASS data used partially overlaps (at $z>0.6$) with the eBOSS+CMASS sample used by \citet{Nadathur:2020b}, and with the combined LOWZ+CMASS sample used here. Our aim here is to provide a coherent analysis of the data superseding that of \citet{Nadathur:2019c}, over a range of redshift bins that can be combined with the eBOSS results of \citet{Nadathur:2020b} without overlap.

\subsubsection{Patchy Mocks}
\label{sec:Patchy Mocks}

The Patchy mocks are a set of 1000 independent mock catalogues on the lightcone created to match the clustering and the survey properties of the BOSS galaxies \citep{Kitaura2016}. They were created using the fast approximate \texttt{Patchy} algorithm \citep{Kitaura2014} based on augmented Lagrangian perturbation theory \citep{Kitaura:2013}, run with fiducial cosmological parameters $\Omega_\mathrm{m}=0.307$, $\Omega_b=0.0482$, $h = 0.6777$, $\sigma_8 = 0.8225$ and $n_s = 0.96$. Mock galaxies were assigned to halos using halo abundance matching, with parameters chosen to reproduce the monopole and quadrupole moments of the galaxy clustering in the BOSS data. The survey mask and selection function were then also matched to those of BOSS.

We use all 1000 of the Patchy mocks both in order to estimate covariance matrices, and a smaller subset of 250 of them to test our analysis method for systematic errors. When doing so we apply the same redshift cuts to the mocks as to the data, i.e. restricting to galaxies with redshift $0.2<z<0.63$ and voids with redshift $0.2<z<0.6$ to avoid redshift overlap with the section of the data included in the eBOSS analysis.

\subsubsection{\uppercase{NSERIES} mocks}
\label{sec:NSERIES Mocks}

The \uppercase{nseries} mocks are a collection of 84 cut-sky mocks made from a 7 independent full $N$-body dark matter simulations with $2048^3$ particles per box and a mass resoluton of $1.5 \times 10^{11} \mathrm{M_\odot}/h$, generated using a flat $\Lambda$CDM cosmology with $\Omega_\mathrm{m} = 0.286$, $\Omega_b = 0.0470$, $h = 0.70$, $\sigma_8 = 0.82$, and $n_s = 0.96$. Halos at redshift $z=0.55$ were populated with mock galaxies using a halo occupation distribution (HOD) prescription adjusted to match the clustering of the CMASS sample. From each of the 7 simulation boxes 12 cut-sky mock catalogues were then created, covering the NGC sky region and matching the selection function for the CMASS sample over the redshift range $0.43<z<0.7$. 

The \uppercase{nseries} mocks do not match the full volume or redshift distribution of the combined BOSS data used in this work, and so cannot be used for estimating covariances. However, unlike the MGS and Patchy mocks, they were created from full $N$-body simulations without approximation and so contain more accurate RSD information on small scales. This makes them useful for testing our analysis pipeline for systematic errors. 

\subsection{eBOSS}
An analysis of the void-galaxy cross-correlation in the eBOSS DR16 LRG sample combined with a portion of the BOSS CMASS sample in the redshift range $0.6<z<1$ was presented by \citet{Nadathur:2020b}. As noted above, we have cut the BOSS galaxy catalogue data to exclude the high-redshift section that was included together with the eBOSS LRGs in that work. Our aim here is provide a consistent voids analysis of all the data over in the MGS, BOSS and eBOSS samples over $0.07<z<1.0$. Since the method used in this work is very similar to that already presented by \citet{Nadathur:2020b} we do not repeat it and so do not directly use the eBOSS data here. Nevertheless in Section~\ref{sec:result} we report the results of this earlier work alongside the new results from the lower redshift samples, and so for completeness alongside MGS and BOSS we also show the eBOSS LRG survey footprint in Figure~\ref{fig:footprints} and the redshift distribution of the eBOSS voids in Figure~\ref{fig:nofz}.

\subsection{Big MultiDark Mocks}
\label{sec:BigMD Mocks}

In order to create templates for functions used in the void modelling described in Section~\ref{sec:model} below, we require access to dark matter information from simulations, which is not available for the MGS, Patchy or \uppercase{nseries} mocks described above. For this we use the Big MultiDark (BigMD) simulation, which is a full $N$-body simulation of 3840$^\mathrm{3}$ dark matter particles evolved in a $2.5 h^{-1} \mathrm{Gpc}$ box using the same cosmology as for the Patchy mocks \citep{Klypin:2020}. We extract halo catalogues from simulation snapshots at $z=0.1,\ z=0.32$, and $z=0.52$ and populate them with mocks according to an HOD prescription to match the clustering of the galaxy data in the different samples. We use HOD parameters matching those of the `Main2' mocks of \citet{Nadathur:2015} at $z=0.1$ to mimic the MGS sample, matching those from \citet{Manera:2015} at $z=0.32$ to mimic the LOWZ sample, and matching those from \citet{Manera:2013} at $z=0.52$ to mimic the CMASS sample. We then cut out sections of the box to match the survey mask in each case, and downsample to match the survey selection functions. We refer to this collection of mock catalogues as the BigMD mocks.

We use the BigMD mocks in order to create templates used in the later analysis only. To do this, we run the reconstruction and void-finding pipeline described in Sections~\ref{sec:recon} and \ref{sec:voidfinding} on these mocks exactly as for the corresponding MGS and BOSS data samples. For the voids thus obtained, we measure the stacked profiles for the enclosed matter density around voids, $\Delta(r, z)$, and the velocity dispersion, $\sigma_{v_{||}}(r, z)$, from the simulation. These functions are used as templates in the modelling as described in Section~\ref{sec:model}. It is worth noting that the steps taken above to match the survey masks in the BigMD mocks are very important, since survey edges can have strong effects on the distribution of void sizes obtained using our algorithm and would thus result in changes to these template functions as well.

\subsection{Fiducial Cosmology}
\label{sec:fiducial cosmo}

When analysing the BOSS data and the BOSS mocks, unless otherwise specified we adopt a reference fiducial cosmological model with $\Omega_\mathrm{m} = 0.307$, $\Omega_\Lambda = 0.693$, $h = 0.676$ and zero curvature in order to convert galaxy redshifts to distances. When analysing the MGS data and the MGS mocks, unless otherwise specified we adopt a reference fiducial cosmological model with $\Omega_\mathrm{m} = 0.31$, $\Omega_\Lambda = 0.69$, $h = 0.676$ and zero curvature in order to convert galaxy redshifts to distances. This cosmological model is very close to that indicated by the \citet{Planck2020} CMB results and matches the cosmology of the Patchy mocks. In Section~\ref{sec:cosmology errors} we assess the dependence of our results on the choice of this fiducial model.

\section{Void Analysis}
\label{sec:analysis}

\subsection{Reconstruction}
\label{sec:recon}
In order to obtain a suitable population of voids for unbiased parameter estimation from the void-galaxy correlation function, we first approximately remove the RSD in the galaxy distribution through the use of reconstruction before applying the void-finding algorithm. This procedure was first advocated by \citet{Nadathur:2019b}, who noted that void-finding performed directly on the redshift-space galaxy distribution leads to samples that violate several of the key assumptions necessary to derive theoretical models of the void-galaxy correlation \citep{Nadathur:2019b, Chuang2017}. These include the assumption that the void-galaxy correlation is spherically symmetric in real space (i.e., that the stack of a large number of voids is spherical on average) and that the mean velocity outflow around void centres is radially directed with a spherically symmetric profile. If void-finding is performed directly in redshift-space, the probability of finding a void becomes dependent on its orientation with respect to the line-of-sight: underdensities aligned along the line-of-sight direction have a higher velocity outflow along that direction, so when viewed in redshift-space appear more strongly stretched and thus appear to have a lower galaxy density at the centre, making them more likely to be selected in the void sample. A recent thorough study by \citet{Correa:2022} characterised this selection effect as an intrinsic ellipticity of samples of redshift-space voids. Such an intrinsic ellipticity is currently not modelled in any theoretical description of the void-galaxy correlation.

To avoid this problem, we instead attempt to remove the selection effect from our sample of voids by first recovering the galaxy field with RSD effects approximately removed. To achieve this, we use the Zeldovich reconstruction algorithm described by \citet{Nadathur:2019b, Nadathur:2019c, Nadathur:2020b} and implemented in the public \texttt{Revolver} code.\footnote{\url{https://github.com/seshnadathur/Revolver}} This uses the iterative fast Fourier transform (FFT) method of \citet{Burden:2015} to solve the Zeldovich equation in redshift space \citep{Zeldovich:1970, Nusser:1994},
\begin{equation}
    \label{eq:Zeldovich}
    \nabla\cdot\mathbf{\Psi}+\frac{f}{b}\nabla\cdot(\mathbf{\Psi}\cdot\mathbf{\hat{r}})\mathbf{\hat{r}} = -\frac{\delta_{g}}{b},
\end{equation}
for the Lagrangian displacement field $\mathbf{\Psi}$, where $f$ is the growth rate, $b$ is the linear galaxy bias, and $\delta_g$ is the galaxy overdensity in redshift space. This step is performed on a $512^3$ grid, and densities estimated on the grid are first smoothed with a Gaussian kernel of width $R_s=10\,h^{-1}$Mpc before solving for the displacement. We then shift individual galaxies by $-\mathbf{\Psi}_\mathrm{RSD}=f\left(\mathbf{\Psi}\cdot \hat{\mathbf{r}}\right)\hat{\mathbf{r}}$ to obtain their (approximate) real-space positions. The parameters $f$ and $b$ are provided as inputs to the reconstruction code, but the results of procedure depend only on $\beta\equiv f/b$.

\begin{figure}
    \centering
    \includegraphics[width=0.48\textwidth]{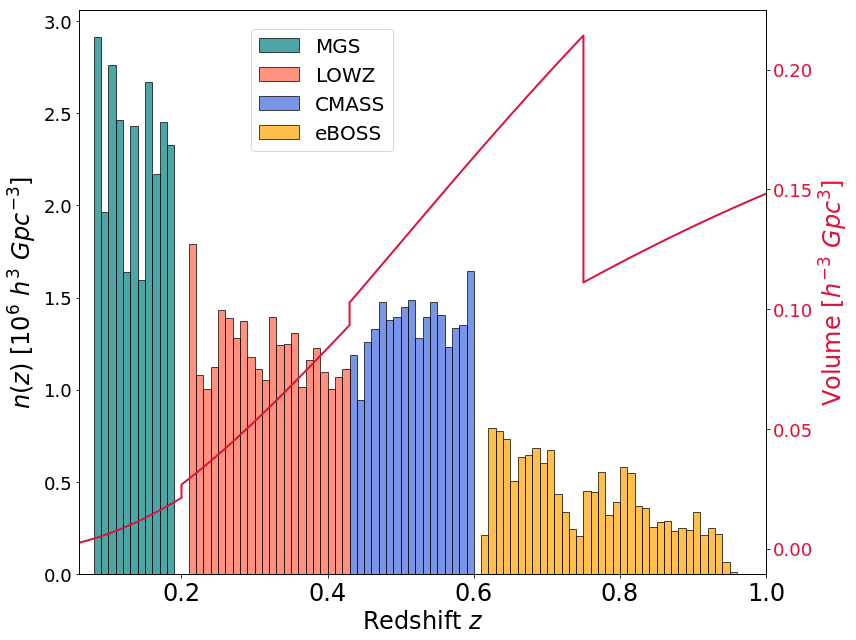}
    \caption{The number density of voids obtained after the application of all selection cuts, across the redshift range of the data (coloured histograms, left axis). Voids from MGS are shown in teal, and those from the combined BOSS LOWZ and CMASS catalogues at $z<0.6$ in orange and blue, where we use the change of colour to indicate the dominant contributing sample to the combination. The yellow histogram shows the $n(z)$ for voids from the combination of the eBOSS LRG and BOSS CMASS catalogues at $z>0.6$ analysed separately by \citet{Nadathur:2020b}, and labelled as `eBOSS' for convenience. Voids were obtained after reconstruction using the fiducial values $\beta=0.31$ (MGS), $\beta=0.37$ (BOSS LOWZ and CMASS) and $\beta=0.35$ (eBOSS). The survey volume in $\Delta z=0.01$ slices is shown as a function of redshift by the red line (right axis), showing the shifts due to changes in the survey mask. 
    }
    \label{fig:nofz} 
\end{figure}

\subsection{Void finding}
\label{sec:voidfinding}
We perform void-finding on the RSD-removed galaxy distribution obtained after the reconstruction step above, using the \texttt{Revolver} code. \texttt{Revolver}  provides several options for the void-finding algorithm; we choose the one based on an adaptation of the \texttt{ZOBOV} void-finder \citep{Neyrinck:2008}. This algorithm first estimates the local density using a Voronoi tessellation of the discrete galaxy distribution, and then identifies voids at the locations of local minima of this density field, using a watershed algorithm to distinguish neighbouring voids. Technical details of the application of \texttt{ZOBOV} and \texttt{Revolver} to survey data, including normalizing density estimates for the survey selection function, the use of systematic weights, the use of buffer particles to limit the tessellation to the observed volume have been provided in several previous publications (see \citealt{Nadathur:2014a} and \citealt{Nadathur:2016a} in particular).

Instead of running reconstruction and then void-finding on the BOSS LOWZ and CMASS catalogues separately, we work directly with the combined BOSS sample composed of both catalogues, including a small redshift range around $z=0.43$ where they overlap. As mentioned in Section \ref{sec:boss}, to avoid duplication of the $z>0.6$ section of the CMASS data that was included with the eBOSS LRGs already analysed by \citet{Nadathur:2020b} we use only the BOSS data below this redshift. However, to minimise the loss of data due to the truncation of voids close to survey boundaries, we work with a slightly larger selection of BOSS galaxies over $0.2<z<0.63$ in the first instance, restricting the final selection to only those voids with centres in the desired $0.2<z<0.6$ redshift range in post-processing.

Working with the combined BOSS sample means that we need to account for the change in the survey footprint between LOWZ and CMASS (see Figure~\ref{fig:footprints}). To do this we modified the standard \texttt{Revolver} algorithm to introduce a layer of buffer particles at $z=0.43$ around the regions in the CMASS footprint that are not included in LOWZ, in the same way as buffers are used around all other survey boundaries. This prevents the tessellation from leaking out of the survey volume and thus guards against recovering spuriously low densities near these boundaries. Additional modifications are also needed to \texttt{Revolver} to correctly calculate the composite survey volume when estimating the local mean galaxy density used to normalise the measured fluctuations. This procedure is the same as that used by \citet{Nadathur:2020b} when analysing the combined eBOSS+CMASS data, which also had a change in survey footprint. No such modifications are required for the MGS catalogue since we run reconstruction and void-finding on this separately without combining with BOSS.

After density field estimation, the individual voids obtained from the watershed algorithm are irregularly shaped and lack spherical symmetry, so the definition of the void ``centre" is not unique. We use the definition introduced by \citet{Nadathur:2015}, which corresponds to the centre of the largest completely empty sphere that can be inscribed within the void and which generally produces a better match to the modelled void-galaxy correlation \citep{Nadathur:2019a}. This is because the validity of the model assumed for the outflow velocity (see equation~\ref{eq:vr} in Section~\ref{sec:model}) can be a less good description of the mean outflow around other centre definitions, resulting in worse overall predictions for $\xi^s$. Finally, following previous works, we apply a minimum size cut to the raw void catalogues, keeping only voids larger than the median obtained size in that catalogue for the final analysis. After this cut is applied we make no further size-based distinctions and treat all voids together in the following. Figure~\ref{fig:nofz} summarises the redshift distribution of the voids obtained from each of the different datasets, after these selection cuts. For completeness we also show the distribution of voids found in the eBOSS+CMASS LRG sample by \citet{Nadathur:2020b}, although we do not repeat the analysis of this data here.

All of the above steps are performed in exactly the same manner on the respective MGS, Patchy and BigMD mock catalogues. For the NSERIES mocks the procedure is very slightly different, since these lie in the CMASS NGC footprint only and so the additional steps above to deal with changes in footprint are not required.

Since void-finding is always performed after reconstruction, the resultant void catalogues inherit a dependence on the parameter $\beta$, which we allow for when fitting to the data. For each catalogue, we perform reconstruction at many values of $\beta$ over a wide range (Section~\ref{sec:likelihood}) and find that the total void numbers obtained vary by up to around $\pm2\%$ with changes in $\beta$. The void numbers shown in Figure~\ref{fig:nofz} and recorded in Table~\ref{tab:zeff} correspond to the values $\beta=0.31$ for MGS, $\beta=0.37$ for BOSS LOWZ and CMASS, and $\beta=0.35$ for eBOSS.

\subsection{Choice of redshift bins}
\label{sec:zbins}

\begin{table*}
    \centering
    \caption{
    Summary of the redshift bins and survey properties for the data used in this work. We show the redshift range, effective redshift, sky area, the total number of voids $N_\mathrm{voids}$, the void size cut applied $R_\mathrm{cut}$, and the number of voids remaining after this cut, $N^\mathrm{cut}_\mathrm{voids}$. The void size cut varies up to $\pm 3\%$ depending on the choice of mock/data galaxy catalogue as well as reconstruction parameter $\beta$. Void numbers vary by up to $\pm2\%$ depending on reconstruction parameter $\beta$, numbers reported are for the fiducial choices of $\beta$. Where a redshift bin is composed of two overlapping samples, one name is chosen to best represent the composite. Data for the final row, $0.6<z<1.0$, are taken from \citet{Nadathur:2020b} and shown here for completeness.
    }
    \label{tab:zeff}
    \begin{tabularx}{0.67\textwidth}{ccccccc}
    \hline 
    \parnoteclear
    \rule{0pt}{2.5ex} Redshift range & Sample name  & Sky area (deg$^2$) & $z_{\mathrm{eff}}$ & $N_{\mathrm{voids}}$ & $R_\mathrm{cut} (h^{-1}\mathrm{Mpc})$  & $N^\mathrm{cut}_{\mathrm{voids}}$ \\[1mm] \hline
    $0.07 < z < 0.2$ & MGS & $6813$ & $0.15$ & $517$ & $40$ & 258       \\
    $0.2 < z < 0.3$  & BOSS LOWZ & $8337$  & $0.26$ & $1009$ & $47$ & 504       \\
    $0.3 < z < 0.4$  & BOSS LOWZ & $8337$ & $0.35$ & $1596$ & $48$ & 798        \\
    $0.4 < z < 0.5$  & BOSS CMASS & $8337$\parnote{for the LOWZ footprint ($0.2 < z < 0.43$)}/$9376$\parnote{for the CMASS footprint ($0.43 < z < 0.75$)} & $0.47$ & $2526$ & $48$ & 1263          \\
    $0.5 < z < 0.6$  & BOSS CMASS & $9376$ & $0.54$ & $3830$ & $49$ & 1915        \\
    $0.6 < z < 1.0$  & eBOSS LRG & $9376^\mathrm{ii}$/$4242$\parnote{for the eBOSS footprint ($z > 0.75$)} & $0.69$ & $4706$ & $49$ & 2341       \\ \hline
    \end{tabularx}
    \parnotes
\end{table*}

In the following, we analyse the MGS data and mocks in a single self-contained redshift bin, $0.07<z<0.2$. However, the BOSS data contain a much larger number of voids extending over larger range of redshifts, so it is possible to split them into a set of narrower redshift bins in order to understand the evolution of the void-galaxy cross-correlation with redshift. Doing so also allows for the fact that the growth rate and galaxy bias, and thus the parameter $\beta$ entering into reconstruction, may evolve with redshift. These considerations lead us to split the voids obtained from the combined BOSS sample into 4 non-overlapping redshift bins: $0.2 < z < 0.3$, $0.3 < z < 0.4$, $0.4 < z < 0.5$, and $0.5 < z < 0.6$. Previous work from \citet{Nadathur:2020b} used the eBOSS+CMASS LRG sample with $0.6<z<1.0$: although we do not re-analyse this data, we report their results again here together with our own. In total therefore we have six redshift bins covering the entire range $0.07<z<1.0$, as summarised in Table~\ref{tab:zeff}.

Within each bin, the effective redshift of the void-galaxy measurement is calculated as a weighted sum
\begin{equation}
    z_\mathrm{eff} = \dfrac{\Sigma_{ij} \left( \frac{Z_i + z_j}{2} w_j \right)}{\Sigma_{ij} w_j},
\end{equation}
where $Z_i$ is the redshift of the void centre, $z_j$ is the galaxy redshift, $w_j$ is the associated galaxy systematic weight, and the sum extends over all void-galaxy pairs up to the maximum separation considered, $s=120\,h^{-1}$Mpc. The effective redshifts for the bins are shown in Table~\ref{tab:zeff}. Where the data covers both galactic caps there was no difference seen between the $z_\mathrm{eff}$ values obtained from the NGC and SGC samples in any redshift bin, so the values reported are for both caps taken together. 

\subsection{Correlation function measurement}

We measure the binned void-galaxy correlation function $\xi^s$ (or $\xi^r$) in redshift space (real-space) in 30 bins of the observed void galaxy separation distance $s$ ($r$) and 80 bins of the cosine of the angle $\mu_s$ ($\mu_r$) between the separation vector and the line-of-sight direction to the void centre using the Landy-Szalay estimator \citep{Landy:1993}:
\begin{equation}
    \label{eq:LSestimator}
    \xi^s(s,\mu_s) = \frac{D_1D_2-D_1R_2-D_2R_1+R_1R_2}{R_1R_2},
\end{equation}
where each term $XY$ refers to the number of pairs for the given populations in the bin, normalized by the effective total number of such pairs. 
Here $D_1$ refers to the void centre positions, $D_2$ to the galaxies, and $R_1$ and $R_2$ to the corresponding sets of unclustered random points matching the angular and redshift distributions and systematic effects of the void and galaxy catalogues but a factor of 50 times larger to minimize shot noise. 
The galaxy randoms $R_2$ are taken from the publicly provided random catalogues for each sample by SDSS. We construct the appropriate void random catalogues ourselves. 

The distribution of voids in the survey volume differs from the distribution of galaxies. This is partly due to an exclusion effect, where voids near survey boundaries are removed due to the possibility of contamination of the tessellation \citep{Nadathur:2016a}. In addition, as the galaxy selection function is controlled for in constructing density estimates, the redshift distribution of voids differs from that of the galaxies. To account for these effects, we construct separate unclustered void random catalogues to match the spatial distribution of the voids, by running the reconstruction and void-finding steps described above on each of the 250 mock MGS or Patchy galaxy mocks respectively to create 250 realisations of the void catalogues. We then randomly draw void positions from the 250 mock catalogues stacked together to make a void random catalogue that has $50$ times more objects than obtained in the survey data. This will result in a catalogue that is very close to being truly random due to the large number of mock catalogues used to generate it.

As described by \citet{Nadathur:2020b}, in computing the pair counts, galaxies and galaxy randoms are weighted by the systematic weights provided in the public data releases. Since these weights have already been accounted for in void-finding, voids and void randoms are all given equal unit weights. 
Where survey data spans two galactic caps, we combine them in the correlation estimation by adding pair counts across caps in Eq.~\ref{eq:LSestimator},  
having first checked that there are no significant systematic offsets between the estimates in each cap.\footnote{We have found that a comparison of results across galactic caps can be useful as a diagnostic of subtle bugs in the reconstruction or void-finding procedure.} We then decompose the measured correlation functions into their Legendre multipole moments, of which we focus here on the monopole, $\xi^s_0(s)$, and quadrupole, $\xi^s_2(s)$, in redshift space, and monopole $\xi^r_0(r)$ in real-space.

Note that we only identify voids in the approximation to the real-space galaxy field obtained from RSD removal after reconstruction. All measured void-galaxy cross-correlations use these void centres and thus implicitly depend on the parameter $\beta$ used in reconstruction. Our measurement of $\xi^s(s, \mu_s)$ uses the original (redshift-space) galaxy positions, but still retains the implicit dependence on $\beta$ from the void identification step. On the other hand, as we do not know the true real-space positions of galaxies we cannot directly determine the true real-space cross-correlation $\xi^r$, and instead estimate it by measuring the cross-correlation with the post- reconstruction galaxy positions with RSD approximately removed. In the following, where necessary we use $\hat\xi^r$ to distinguish this measured estimate of the true real-space cross-correlation $\xi^r$.

\subsection{Model}
\label{sec:model}

In the absence of Alcock-Paczynski distortions, the redshift-space void-galaxy cross-correlation function $\xi^s(\mathbf{s})$ is related to  the real-space version $\xi^r(\mathbf{r})$ by 
\begin{equation}
    \label{eq:streaming}
    1 + \xi^s(s_\perp, s_{||}) = \int_{-\infty}^{\infty}\left(1 + \xi^r(\mathbf{r})\right)P(v_{||}, \mathbf{r})\,\rmn{d}v_{||},
\end{equation}
where $P(v_{||}, \mathbf{r})$ is the position-dependent PDF of galaxy velocities parallel to the line of sight direction, $v_{||}$, and the real-space void-galaxy separation vector $\mathbf{r}$ and its redshift-space equivalent $\mathbf{s}$ have components perpendicular to and parallel to the line of sight direction that are related by $s_\perp = r_\perp$ and
\begin{equation}
    \label{eq:coords}
    s_{||} = r_{||} + \frac{v_{||}}{aH},
\end{equation}
respectively, where $a$ is the scale factor and $H$ the Hubble rate at the redshift of the void. This expression is general and exact if the number of void-galaxy pairs is conserved. Although the number of voids is \emph{not} conserved under the application of void-finding separately to real- and redshift-space galaxy distributions \citep{Chuang2017, Nadathur:2019b, Correa:2022}, the assumption of pair conservation holds by construction in our case, since we use the same sample of voids (in our case identified in the reconstructed galaxy field) for evaluation of both $\xi^r(\mathbf{r})$ and $\xi^s(\mathbf{s})$.

When considering the distribution of galaxies around a stack of voids, we can further assume spherical symmetry in real-space, which means that $\xi^r(\mathbf{r})=\xi^r(r)$, and the velocity distribution $P(v_{||}, \mathbf{r})$ at each $r$ is symmetric around the mean value $v_r(r)\mu_r$, where $\mathbf{v}(r) = v_r(r)\hat{\mathbf{r}}$ is the (radially directed) coherent mean galaxy outflow velocity around the void and $\mu_r=r_{||}/r=\cos\theta$ where $\theta$ is the angle between the void-galaxy separation vector and the line-of-sight. If we introduce a change of variables $\tilde{v}=v_{||}-v_r(r)\mu_r$, then by using the relations
\begin{eqnarray}
    \label{eq:dv}
    \diff{\tilde{v}}{v_{||}}&=&1 - r_{||}\dv{v_{||}}\left(\frac{v_r}{r}\right) - \left(\frac{v_r}{r}\right)\dv{r_{||}}{v_{||}}\,,\\
    \label{eq:dr}
    \dv{r}{v_{||}}&=&\frac{r_{||}}{r}\dv{r_{||}}{v_{||}}\,,\\
    \label{eq:drpar}
    \dv{r_{||}}{v_{||}}&=&-\frac{1}{aH}\,,
\end{eqnarray}
we can rewrite Eq.~\ref{eq:streaming} as
\begin{equation}
    \label{eq:full_model}
    1 + \xi^s(s, \mu_s) = \int\left(1 + \xi^r(r)\right)\left[1+\frac{v_r}{raH} + \frac{rv_r^\prime-v_r}{raH}\mu_r^2\right]^{-1}P(\tilde{v},r)\rmn{d}\tilde{v}\,,
\end{equation}
where $\mu_s=s_{||}/s$, the term in the square brackets is $\dv{\tilde{v}}{v_{||}}$, and $'$ denotes the derivative with respect to $r$. The term $P(\tilde{v},r)$ now represents incoherent dispersion as we have explicitly removed the coherent outflow from the velocity. Eq.~\ref{eq:full_model} is exactly the model derived by \citet{Nadathur:2019a}, who derived it using the Jacobian of the mapping between $\mathbf{s}$ and $\mathbf{r}$ and then added in a Gaussian dispersion term $P(\tilde{v},r)$, which was required to fit the simulations. As shown in the derivation above, this term naturally arises in the streaming model as the incoherent component of $P(v_{||}, \mathbf{r})$ in Eq.~\ref{eq:streaming}.

Note that the key assumption of spherical symmetry in real-space requires \emph{both} statistical isotropy of the Universe as a whole \emph{and} that void selection also maintains statistical isotropy -- i.e., that the process of identification of voids has no orientation-dependent bias. This cannot in principle be true if void-finding is applied to the redshift-space galaxy density field, which already contains line-of-sight anisotropies due to RSD. In this case underdensities with larger outflow velocities along the line-of-sight are preferentially selected as voids, and this selection bias means that neither $\xi^r(\mathbf{r})$ nor $P(v_{||}, \mathbf{r})$ are isotropic, and that the PDF is not symmetric about the mean. This leads to a large additional contribution to $\xi^s$ \citep{Nadathur:2019b, Correa:2022} that cannot currently be modelled. It is precisely to remove this orientation-dependent void selection that we employ the additional reconstruction step in our observational pipeline (Section \ref{sec:recon}).

When this is done, it has been shown empirically from comparison with simulations that the PDF $P(\tilde{v}, r)$ is close to Gaussian over a range of scales \citep{Nadathur:2019a, Paillas:2021}. Deviations from Gaussianity occur at large $r$, where the effect of convolution with $P(\tilde{v}, r)$ in Eq.~\ref{eq:full_model} is itself negligible. We therefore assume a zero-mean Gaussian PDF with standard deviation $\sigma_{v_{||}}$,
\begin{equation}
    \label{eq:pdf}
    P(\tilde{v},r) = \frac{1}{\sqrt{2\pi}\sigma_{v_\parallel}(r)} \exp\left(-\frac{\tilde{v}^2}{2\sigma_{v_\parallel}^2(r)} \right)\,.
\end{equation}
Evaluation of Eq.~\ref{eq:full_model} then gives very similar results to the Gaussian streaming model (GSM) that has also been used for similar cross-correlation analyses \citep{Paz:2013,Cai:2016a,Paillas:2021}.\footnote{In previous work \citep[e.g.][]{Nadathur:2020b}, we erroneously stated that Eq.~\ref{eq:full_model} and the GSM produced numerically significantly different results. This was due to a bug in our implementation of the GSM, though evaluation of Eq.~\ref{eq:full_model} was unaffected. We thank Enrique Paillas for helping resolve the issue. Both these models, as well as some others, are now implemented in the public \texttt{Victor} package.}

In order to use Eq.~\ref{eq:full_model}, we still need to specify a model for the mean coherent outflow velocity $v_r(r)$. Results from simulations \citep[e.g.,][]{Hamaus:2014a,Nadathur:2019a,Nadathur:2019c} show that for voids similar to those used in this analysis, the result obtained from linear perturbation theory applied to the continuity equation,
\begin{equation}
    \label{eq:vr}
    v_r(r) = -\frac{1}{3}faHr\Delta(r)\,,
\end{equation}
where $f$ is the linear growth rate and $\Delta(r)$ is the average mass density contrast within radius $r$ of the void centre,
\begin{equation}
    \label{eq:Delta def}
    \Delta(r) = \frac{3}{r^3}\int_0^r\delta(y)y^2\,\rmn{d}y\,.
\end{equation} 
provides a good description of the outflow velocity. 
However, the validity of Eq.~\ref{eq:vr} depends on the choice of void-finding algorithm and the use of alternative algorithms can lead to agreement that is not as good, requiring additional corrections to Eq.~\ref{eq:vr} \citep{Paillas:2021}. Although Eq.~\ref{eq:vr} is nominally obtained from linear perturbation theory,\footnote{But note that the enclosed mass density profile $\Delta(r)$ here is \emph{not} the linear theory prediction, but rather the fully non-linear density that would be measured in simulations.} we do not make further approximations of linearity in our analysis. In particular, Eq. ~\ref{eq:full_model} is evaluated directly, without expanding in powers of $\Delta$ as is sometimes done.

While Eq.~\ref{eq:vr} specifies the form of the dependence on the growth rate $f$, it still refers to the void matter density profile $\Delta(r)$, which is in principle unknown. Some works \citep[e.g.][]{Hamaus:2017a,Hawken:2020,Aubert20a} model this term using a simple linear bias prescription, $\Delta(r)=\xi^r(r)/b$, where $b$ is the large-scale linear galaxy bias, to relate it to the real-space void-galaxy correlation (which can in turn be directly measured from the data, where necessary). However we have found that this assumption is often a poor approximation and can lead to strongly biased parameter estimates \citep{Nadathur:2019a,Nadathur:2020b}. Therefore we follow a \emph{template-fitting} approach instead. We calibrate a fiducial template $\Delta^\rmn{fid}(r)$ using galaxy voids and dark matter information in the BigMD simulation at snapshot redshift $z_\rmn{ref}$, and allow the amplitude of this template profile to scale freely with the parameter $\sigma_8$ describing the amplitude of matter perturbations:
\begin{equation}
    \label{eq:Delta}
    \Delta(r; z) = \frac{\sigma_8(z)}{\sigma_8^\mathrm{BigMD}(z_\rmn{ref})}\Delta^\mathrm{fid}(r; z_\rmn{ref})\,.
\end{equation}
This linear scaling of $\Delta$ with $\sigma_8$ was verified through comparison with simulations constructed with differing $\sigma_8$ by \citet{Nadathur:2019c}. We construct template profiles $\Delta^\mathrm{fid}(r; z_\rmn{ref})$ from snapshots at redshifts $z_\rmn{ref}=0.1, 0.32, 0.52$, and use the closest one to the redshift bin in question.

In a similar spirit, we do not model the dispersion function $\sigma_{v_{||}}(r)$ but instead follow \citet{Nadathur:2019c} by also constructing templates for this function from the BigMD simulation, and allowing the amplitude of this template, denoted by $\sigma_v$ and corresponding to the asymptotic value of $\sigma_{v_{||}}(r)$ at large $r$, to be a free parameter in the model fits. 

Even once $\Delta(r)$ and $\sigma_{v_{||}}(r)$ have been specified in this way, Eq.~\ref{eq:full_model} only describes the relationship or mapping between the real-space and redshift-space correlation functions $\xi^r$ and $\xi^s$. We do not attempt to describe $\xi^r$ itself from first principles, since this would at a minimum require a mathematical model of the action of the void-finding algorithm in addition to cosmological theory. Instead we follow \citet{Nadathur:2019c} and \citet{Nadathur:2020b} by using the estimate $\langle\hat\xi^r\rangle$ determined from the 250 MGS or Patchy mocks instead, where $\hat\xi^r$ is the measured void-galaxy correlation obtained using the RSD-removed mock galaxy field after reconstruction, and $\langle\rangle$ denotes the average over all the mocks. 

An alternative to using this average over the mocks could be to use the estimate $\hat\xi^r(r)$ obtained directly from measurement in the SDSS data itself -- this would be analogous to the approach taken by \citet{Hamaus:2021} to approximate $\xi^r(r)$ from the data, except that they used a deprojection technique while we use reconstruction to accomplish the RSD removal. Such an approach has two potential disadvantages, however: the estimate of $\xi^r(r)$, being derived from only a single realisation rather than the mean of 250, is significantly noisier; and this noise is significantly correlated with measurement noise in $\xi^s(\mathbf{s})$, since both are measured from the same data. This introduces a significant correlation between the model prediction and the data vector to which it is being compared. This correlation would need to be carefully accounted for in the covariance matrix and propagated through the likelihood -- if this is not done, we find that the fit to the data returns an artificially low $\chi^2$ and can lead to a systematic bias in the recovered cosmological parameters. In contrast, for the procedure we use here such accounting is not necessary, since the mean $\langle\hat\xi^r\rangle$ over the mocks cannot be correlated with $\xi^s$ measured in the SDSS data.

While Eq.~\ref{eq:full_model} is only valid in the true cosmology without Alcock-Paczynski (AP) distortions \citep{Alcock:1979}, it is simple to extend this to accommodate differences arising due to choice of the fiducial model used to convert observed redshifts to distances. We define the $\alpha$ scaling parameters
\begin{equation}
    \label{eq:alphas}
    \aperp\equiv\frac{D_\mathrm{M}(z)}{D_\mathrm{M}^\mathrm{fid}(z)}\;;\;\apar\equiv\frac{D_\mathrm{H}(z)}{D_\mathrm{H}^\mathrm{fid}(z)}\,,
\end{equation}
where $D_\mathrm{M}(z)$ is the comoving angular diameter distance and $D_\mathrm{H}(z) = c/H(z)$ is the Hubble distance at redshift $z$, and then 
\begin{equation}
    \label{eq:model_with_alphas}
    \xi^s(s_\perp, s_\parallel) = \xi^{s,\mathrm{fid}}\left(\aperp s_\perp^\mathrm{fid}, \apar s_\parallel^\mathrm{fid}\right)\,,
\end{equation}
where the superscript $^\rmn{fid}$ indicates quantities in the fiducial cosmological model. In calculations using Eq.~\ref{eq:full_model}, we always rescale the input functions $\hat\xi^r(r)$, $\Delta(r)$ and $\sigma_{v_\parallel}(r)$ with the AP $\alpha$ parameters as described by \citet{Nadathur:2019c}, equivalent to changing the apparent void size by $r\rightarrow\aperp^{2/3}\apar^{1/3}r$ to account for AP dilations. This means we do not use the absolute void size as a standard ruler and so the model prediction is sensitive only to the ratio $\aperp/\apar$. 

All model calculations are made using the public Python package \texttt{Victor}.\footnote{\url{https://github.com/seshnadathur/victor}} In addition to the model of Eq.~\ref{eq:full_model} described here, \texttt{Victor} also implements the GSM and a number of other models that have been used in the literature in order to enable easy comparison of theoretical approaches.

Finally, we note again that as with the measured data vector, the model prediction for $\xi^s$ inherits an implicit dependence on $\beta=f/b$ through $\xi^r(r)$ which is determined using reconstructed galaxy data. This is accommodated as described in Section~\ref{sec:likelihood} below.

\subsection{Likelihood}
\label{sec:likelihood}

We perform all comparisons of model and data using the \texttt{Victor} void-galaxy correlation analysis tool, with data vector $\boldsymbol{\xi}^s=\left(\xi^s_0(s),\xi^s_2(s)\right)$ formed from the monopole and quadrupole moments of the redshift-space correlation function. It is simple to add the hexadecapole and other higher order moments to the analysis in \texttt{Victor}, but they do not add much information at the measurement precision of current data and so are ignored in what follows. The calculation of the theory model $\boldsymbol{\xi}^{s,\rmn{th}}$ sketched above has an explicit dependence on $f\sigma_8$, $\aperp/\apar$ and $\sigma_v$, and an implicit dependence on $\beta$, while the measured data vector $\mathbf{\xi}^s$ also depends on $\beta$. The parameter space we explore is therefore 4-dimensional.

We use the mocks to construct an estimate of the covariance matrix,
\begin{equation}
    \label{eq:covmat}
    \mathbf{C} = \frac{1}{n_s-1}\sum_{k=1}^{n_s}\left(\boldsymbol{{\xi}^s}^k-\overline{\boldsymbol{{\xi}^s}^k}\right)\left(\boldsymbol{{\xi}^s}^k-\overline{\boldsymbol{{\xi}^s}^k}\right)\,,
\end{equation}
from the $n_s=1000$ mocks, 
where $\boldsymbol{{\xi}^s}^k$ is the measured data vector in the $k$th mock realisation, and $\overline{\boldsymbol{{\xi}^s}^k}$ is the mean over the mocks. At a given point $\left(f\sigma_8, \beta, \aperp/\apar, \sigma_v\right)$ in parameter space, we then use this covariance matrix estimate to obtain the $\chi^2$ for the model fit,
\begin{equation}
    \label{eq:chi2}
    \chi^2 = \left(\boldsymbol{\xi}^{s,\mathrm{th}}-\boldsymbol{\xi}^s\right) \mathbf{C}^{-1} \left(\boldsymbol{\xi}^{s,\mathrm{th}}-\boldsymbol{\xi}^s\right)\,.
\end{equation}
In doing so we treat the covariance matrix as fixed and do not attempt to account for its dependence on cosmology.

Since the estimate of the covariance matrix in Eq.~\ref{eq:covmat} from the mocks is itself uncertain, it is necessary to propagate this uncertainty through to the likelihood. To do this, we use the procedure outlined by \citet{Percival2021} to calculate the posterior 
\begin{equation}
    f(\boldsymbol{\theta}|\boldsymbol{\xi}^s) \propto
    \left[1+\frac{\chi^2}{(n_s-1)}\right]
        ^{-\frac{m}{2}}\,,
\end{equation}
where the power law index $m$ is given by 
\begin{eqnarray}
  m&=&n_\theta+2+
    \frac{n_s-1+B(n_d-n_\theta)}{1+B(n_d-n_\theta)}\,,\\
      B&=&\frac{(n_s-n_d-2)}{(n_s-n_d-1)(n_s-n_d-4)}\,,
\end{eqnarray}
and $n_s=1000$ is the number of mocks, $n_d=60$ the number of data points fitted and $n_\theta=4$ the number of model parameters. This procedure adopts flat and uninformative priors on all parameters and a prior on the covariance matrix such that we can match the Bayesian results to frequentist expectations to first order, allowing us to compare credible intervals derived from the posterior to confidence regions derived from the scatter of results from the mocks \citep{Percival2021}. 

In order to explore the model parameter space, we use the MCMC sampling implemented in \texttt{Victor} via an interface with the \texttt{Cobaya} sampling package \citep{Cobaya:code, Cobaya:paper}. Since directly repeating the reconstruction, void-finding and cross-correlation measurements at each value of $\beta$ would make the MCMC prohibitively expensive, we adopt a time-saving interpolation strategy. Before running the MCMC we compute all the necessary cross-correlations on a closely-spaced grid of $\beta$ values, $\beta_i$. During the MCMC run, for each input $\beta$ we evaluate the likelihood twice at the grid points bracketing it $\beta_i\leq\beta\leq\beta_{i+1}$, and then linearly interpolate between these values. This differs slightly from the method used by \citet{Nadathur:2019c,Nadathur:2020b}, who performed the interpolation at the level of the correlation functions. It was found during this analysis that interpolating at the level of the correlation function results in a reduction of noise in the region in between values of $\beta_i$ and $\beta_{i+1}$. This effect results in a lower $\chi^2$ between the $\beta$ grid points. The choice was made in this work to interpolate between the grid points bracketing the $\beta$ value to alleviate this artificial reduction in $\chi^2$. As a result of this the best fit value of $\beta$ will always be found at a $\beta$ grid point and so a closely spaced grid must be used in the analysis.

\section{Tests of systematic errors}\label{sec:systematic}
\begin{figure*}
    \centering
    \includegraphics[width=0.9\textwidth]{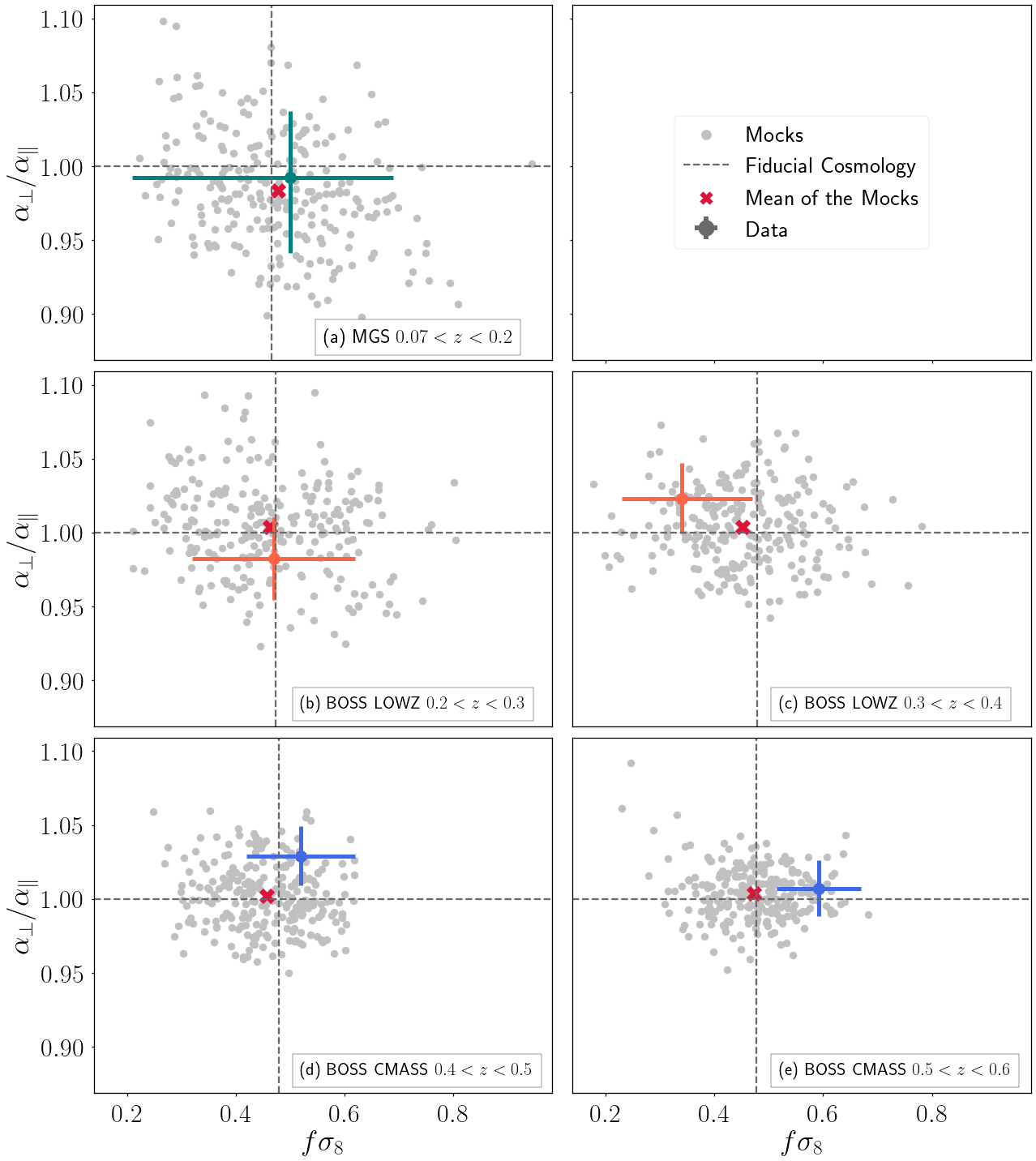}
    \caption{\label{fig:sdss_scatter} Model performance in recovering the fiducial values of $f\sigma_8$ and $\alpha_\perp/\alpha_{||}$ in mocks. Grey points show the results for the mean values of $f\sigma_8$ and $\alpha_\perp/\alpha_{||}$ (after marginalising over other fit parameters) obtained from repeating the analysis on each of 250 MGS mocks ($0.07<z<0.2$ bin) and 250 Patchy mocks (all other redshift bins) when evaluated in their own fiducial cosmology. The means of these individual results are shown by the red crosses, and the expected truth values of the parameters are indicated by the dashed lines. The differences between the means and the expected values are quantified and included in the systematic error budget (Section ~\ref{sec:systematic}). The coloured points with error bars indicate the result and the associated 1$\sigma$ statistical errors obtained from the MGS (teal), BOSS LOWZ (orange) and BOSS CMASS (blue) data in the same redshift bins.
    }
\end{figure*}

In this section, we describe below a series of tests to examine sources of systematic error and quantify their contribution to the total error budget for the two model parameters of cosmological interest, $f\sigma_8$ and $\alpha_\perp/\alpha_{||}$. We divide these into two categories: errors that may be introduced by limitations in the theoretical model when applied to representative survey data (referred to as `modelling systematics'), and errors that can be associated with the choice of the fiducial cosmological model in which the data analysis is performed. We discuss these separately in Sections~\ref{sec:modelling} and \ref{sec:cosmology errors} and combine results into a total systematic error budget in Section~\ref{sec:totalError}.

\subsection{Modelling Systematics}\label{sec:modelling}
\begin{table*}
    \caption{Performance of the model when analyzing mock catalogues. The mocks are analyzed in their own cosmology, indicated in bold, as well as perturbations around this at $\Omega_\mathrm{m}=0.28$ and $\Omega_\mathrm{m}=0.34$ to find the error budget due to analyzing the mocks in the wrong cosmology. The differences are shown between the mean values obtained by the mocks and the expected values. The $2\sigma$ uncertainties on these differences are found as twice the mean of the 1D marginalized parameter uncertainties in the individual mocks multiplied by $1/\sqrt{N_\mathrm{mocks}}$. Values that are more than $2\sigma$ deviant from expectation are highlighted in bold font.}
    \label{tab:systematic-errors}
    \begin{tabular}{cccccccc}
    \hline
    Mock & $N_\mathrm{mocks}$ & Redshift Range & Ref Cosmology & $\langle f \sigma_8 \rangle$ & $\langle \aperp/\apar \rangle$ & $\Delta(f\sigma_8) \pm 2 \sigma$ & $\Delta(\aperp/\apar) \pm 2 \sigma$ \\ \hline
    \multirow{ 3}{*}{MGS} & \multirow{ 3}{*}{250} & \multirow{ 3}{*}{$0.07 < z < 0.2$} 
            & $\Omega_\mathrm{m} = 0.28$             & $0.4661$ & $0.9968$ & \boldmath{$0.0464 \pm 0.0334$} & \boldmath{$-0.0098 \pm 0.0053$}  \\ 
        & & & \boldmath{$\Omega_\mathrm{m} = 0.31$}  & $0.4661$ & $1.0000$ & $0.0112 \pm 0.0252$ & \boldmath{$-0.0167 \pm 0.0055$} \\
        & & & $\Omega_\mathrm{m} = 0.34$             & $0.4661$ & $1.0035$ & $0.0162 \pm 0.0297$ & \boldmath{$-0.0094 \pm 0.0053$} \\[1mm]
        
    \multirow{ 3}{*}{Patchy} & \multirow{ 3}{*}{$250$} & \multirow{ 3}{*}{$0.2 < z < 0.3$}  
            & $\Omega_\mathrm{m} = 0.28$             & $0.4733$ & $0.9945$ & $-0.0097 \pm 0.0208$ & $0.0044 \pm 0.0047$ \\
        & & & \boldmath{$\Omega_\mathrm{m} = 0.307$} & $0.4733$ & $1.0000$ & $-0.0108 \pm 0.0210$ & $0.0034 \pm 0.0047$ \\
        & & & $\Omega_\mathrm{m} = 0.34$             & $0.4733$ & $1.0061$ & $0.0111 \pm 0.0199$ & $-0.0014 \pm 0.0047$ \\[1mm]
        
    \multirow{ 3}{*}{Patchy} & \multirow{ 3}{*}{$250$} & \multirow{ 3}{*}{$0.3 < z < 0.4$}  
            & $\Omega_\mathrm{m} = 0.28$             & $0.4786$ & $0.9928$ & \boldmath{$-0.0275 \pm 0.0153$} & \boldmath{$0.0065 \pm 0.0038$} \\
        & & & \boldmath{$\Omega_\mathrm{m} = 0.307$} & $0.4786$ & $1.0000$ & \boldmath{$-0.0269 \pm 0.0150$} & $0.0038 \pm 0.0039$ \\
        & & & $\Omega_\mathrm{m} = 0.34$             & $0.4786$ & $1.0080$ & \boldmath{$-0.0190 \pm 0.0164$} & $-0.0013 \pm 0.0037$ \\[1mm]
        
    \multirow{ 3}{*}{Patchy} & \multirow{ 3}{*}{$250$} & \multirow{ 3}{*}{$0.4 < z < 0.5$}  
            & $\Omega_\mathrm{m} = 0.28$             & $0.4795$ & $0.9907$ & \boldmath{$-0.0221 \pm 0.0138$} & \boldmath{$0.0094 \pm 0.0029$} \\
        & & & \boldmath{$\Omega_\mathrm{m} = 0.307$} & $0.4795$ & $1.0000$ & \boldmath{$-0.0226 \pm 0.0130$} & $0.0018 \pm 0.0028$ \\
        & & & $\Omega_\mathrm{m} = 0.34$             & $0.4795$ & $1.0102$ & \boldmath{$-0.0271 \pm 0.0116$} & $-0.0001 \pm 0.0029$ \\[1mm]
        
    \multirow{ 3}{*}{Patchy} & \multirow{ 3}{*}{$250$} & \multirow{ 3}{*}{$0.5 < z < 0.6$} 
            & $\Omega_\mathrm{m} = 0.28$             & $0.4773$ & $0.9896$ & $-0.0112 \pm 0.0113$ & \boldmath{$0.0102 \pm 0.0026$} \\
        & & & \boldmath{$\Omega_\mathrm{m} = 0.307$} & $0.4773$ & $1.0000$ & $-0.0047 \pm 0.0111$ & \boldmath{$0.0037 \pm 0.0026$} \\
        & & & $\Omega_\mathrm{m} = 0.34$             & $0.4773$ & $1.0113$ & $-0.0073 \pm 0.0103$ & $-0.0001 \pm 0.0025$ \\[1mm]
        
    NSERIES & $84$ & $0.43 < z < 0.5$ 
            & \boldmath{$\Omega_\mathrm{m} = 0.286$} & $0.4687$ & $1.0000$  & $-0.0065 \pm 0.0246$ & $-0.0053 \pm 0.0056$ \\[1mm]
            
    NSERIES & $84$ & $0.5 < z < 0.6$ 
            & \boldmath{$\Omega_\mathrm{m} = 0.286$} & $0.4687$ & $1.0000$ & $-0.0154 \pm 0.0173$ & $-0.0008 \pm 0.0045$ \\ \hline
            
    \end{tabular}
\end{table*}

We use the mock catalogues described in Section~\ref{sec:data} and run the full measurement and fitting pipeline on 250 realisations each of the MGS and Patchy mocks. These mocks are treated in exactly the same manner as the corresponding MGS and BOSS data samples. This involves running void finding on the whole sample, splitting the voids found into the same redshift bins as the data, and performing subsequent analysis in these bins. We initially use the true cosmology of the mocks as the fiducial model for converting redshifts to distance. From the fits to each mock we obtain the mean values of the cosmologically interesting parameters $f\sigma_8$ and $\alpha_\perp/\alpha_{||}$, after marginalising over $\beta$ and $\sigma_v$, and compare the averages over all mocks, $\langle f\sigma_8\rangle$ and $\langle \alpha_\perp/\alpha_\parallel \rangle$ to the known values for the mock cosmology in that redshift bin. These results are summarised in Table~\ref{tab:systematic-errors}, and Figure~\ref{fig:sdss_scatter} shows the scatter in the recovered values over all 250 mocks in each redshift bin.

We consider a statistically significant systematic error to be detected when the mean over the mocks for a given parameter differs from its expectation value by more than twice the expected statistical error in the mean, calculated as $1/\sqrt{N_\rmn{mocks}}$ times the average marginalised 1D parameter uncertainty for a single mock. We see significant offsets in $f\sigma_8$ for two redshift bins ($0.3 < z < 0.4$ and $0.4 < z < 0.5$) and in $\alpha_\perp/\alpha_\parallel$ for two redshift bins ($0.07 < z < 0.2$ and $0.5 < z < 0.6$). These offsets are however always small compared to the statistical precision that can be obtained in the data. In Section \ref{sec:totalError} below we describe how these are incorporated into the total systematic error budget.

In addition to analysing the MGS and Patchy approximate mocks, in Table~\ref{tab:systematic-errors} we also show the equivalent results obtained from fitting to the 84 NSERIES mocks. These NSERIES mocks cover only a subset of the full redshift range of the BOSS data and only the NGC sky region. We analyse them in modified redshift bins $0.43<z<0.5$ and $0.5<z<0.6$ and using a covariance matrix appropriate to the reduced sky area, but otherwise treat them in the same manner as for the data catalogues. Since they do not match the sky footprint and redshift range of the SDSS data, we do not include offsets determined from the NSERIES mocks in our estimation of the total systematic error budget. Nevertheless, as these mocks are drawn from the full $N$-body simulations they are expected to reproduce the true RSD signal to higher accuracy and smaller scales than possible with the approximate Patchy mocks. It is therefore reassuring that, to within the slightly reduced precision afforded by the smaller number of mock realisations, no systematic offsets are found in the recovered values of either $f\sigma_8$ or $\alpha_\perp/\alpha_{||}$ from NSERIES.

\begin{figure*}
    \centering
    \includegraphics[width=0.8\textwidth]{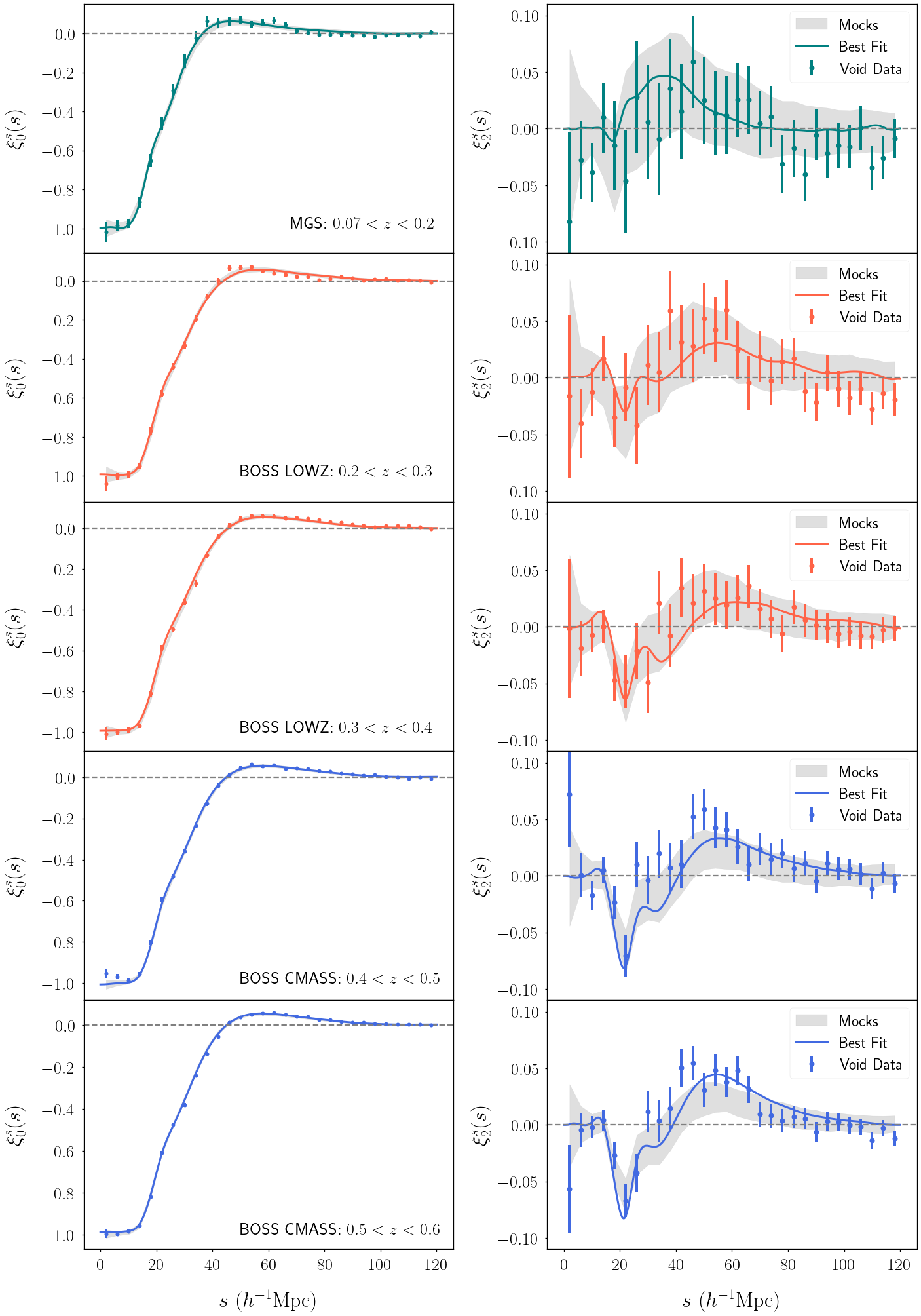}
    \caption{\label{fig:bestfit} 
    Multipole moments of the measured void-galaxy cross-correlation from the MGS and BOSS data. The left column shows the monopole moments and the right column the quadrupole moments, different rows correspond to the data sample and redshift slice indicated. The observed data vector depends on $\beta$; results here are shown for the best-fit $\beta$ values in each redshift bin, $\beta=0.32,0.30,0.41,0.41,0.43$ from top to bottom. Error bars are derived from diagonal entries of the covariance matrix obtained from 1000 realisations of the respective MGS or Patchy mocks. In each panel the solid black line shows the best-fit model of Eq.~\ref{eq:full_model}. The shaded regions show the 68\% confidence range for the same multipole moments measured in the mocks, evaluated at the same values of $\beta$ as the observed data vector.}
\end{figure*}

\subsection{Effect of the fiducial cosmology}
\label{sec:cosmology errors}

We estimate the systematic error introduced by performing the analysis with a fiducial cosmological model that differs from the true cosmology by repeating the entire analysis over the 250 MGS and Patchy mocks using different cosmological
models. We consider perturbations around the true cosmology of the mocks by setting $\Omega_\mathrm{m} = 0.28$ and $\Omega_\mathrm{m} = 0.34$. All other aspects of the analysis remain the same as before. The recovered mean values $\langle f\sigma_8\rangle$ and $\langle \alpha_\perp/\alpha_\parallel \rangle$ over all the mocks are summarized in Table \ref{tab:systematic-errors} for each redshift bin and cosmology tested.

As above, we consider a statistically significant systematic error to be detected if the mean value over the mocks differs from the truth by more than twice the expected error in the mean, estimated as $1/\sqrt{N_\mathrm{mocks}}$ times the average marginalised 1D parameter uncertainty for a single mock. Table \ref{tab:systematic-errors} shows that this threshold is exceeded for several redshift bins when $\Omega^\mathrm{fid}_\mathrm{m}=0.28$, and occasionally when $\Omega^\mathrm{fid}_\mathrm{m}=0.34$, though the differences are still small compared to the statistical uncertainty in fitting to a single realisation. This increased occurrence of systematic offsets may point to a deficiency in the modelling used when the fiducial model is far from the truth. Further improvements to the modelling in future work may be able to eliminate this source of error, but for the current paper we instead incorporate this into the total systematic error budget as described below.

\subsection{Total Systematic Error Budget}\label{sec:totalError}

To determine the total systematic error in our measurements we use the results for the differences with respect to the true values, $\Delta(f\sigma_8)$ and $\Delta(\alpha_\perp/\alpha_{||})$, for the MGS and Patchy mocks shown in Table~\ref{tab:systematic-errors}. These reported offsets do not show strong correlations across redshift bins and choices of fiducial cosmology, comparing both positive and negative values, some of which are statistically significant. We therefore model them as arising from a draw from an underlying distribution, and use the results shown in the table to estimate the mean and standard deviation of this distribution. We compute the weighted means and standard deviations for $\Delta(f\sigma_8)$ and $\Delta(\alpha_\perp/\alpha_{||})$ separately using inverse variance weights for each row in Table~\ref{tab:systematic-errors}, with the variance corresponding to the statistical uncertainty in the measurement of the mean in each redshift bin.

Following this procedure, for $f\sigma_8$ we estimate that the systematic error in the measurement has a mean $\sigma_{\mathrm{sys, offset}} = -0.0113$ and a standard deviation $\sigma_{\mathrm{sys, error}} =  0.0161$. The corresponding values for $\alpha_\perp/\alpha_\parallel$ are $\sigma_{\mathrm{sys, offset}} = 0.0029$ and $\sigma_{\mathrm{sys, error}} = 0.0061$. To incorporate these into the the total systematic error budget, we: 
\begin{enumerate}
    \item subtract the corresponding value of $\sigma_{\mathrm{sys, offset}}$ from the reported mean result for each parameter to correct mean bias, and
    \item add both systematic error estimates in quadrature to the statistical error to determine the total error budget, $\sigma_\mathrm{total} = \sqrt{\sigma_\mathrm{syst, offset}^2 + \sigma_\mathrm{syst, error}^2 + \sigma_\mathrm{stat}^2 }$ (where $\sigma_\mathrm{stat}$ is obtained from the fit to the SDSS data).
\end{enumerate}
The results for each redshift bin are summarised in Table~\ref{tab:total error}.
In each redshift bin $\sigma_{\mathrm{sys, offset}}$ and $\sigma_{\mathrm{sys, error}}$ are both small compared to the statistical error $\sigma_\mathrm{stat}$ so they result in only a modest increase in the total error budget in each case.

\begin{table}
    \caption{Summary of the total error budget for measurement of $f \sigma_8$ and $\alpha_\perp / \alpha_\parallel$ in each redshift bin. Statistical errors $\sigma_{\mathrm{stat}}$ are determined from posterior fits to the data. The total systematic error budget is determined by adding in quadrature the individual contributions described in Section~\ref{sec:totalError}, $\sigma_{\mathrm{total}} = \sqrt{\sigma_{\mathrm{syst, offset}}^2 + \sigma_{\mathrm{syst, error}}^2 + \sigma_{\mathrm{stat}}^2 }$. Here $\sigma_{\mathrm{syst, offset}} = -0.0113$ for $f\sigma_8$ and $0.0029$ for $\alpha_\perp/\alpha_\parallel$, and $\sigma_{\mathrm{syst, error}} = 0.0161$ for $f\sigma_8$ and $0.0061$ for $\alpha_\perp/\alpha_\parallel$.}
    \label{tab:total error}
    \begin{center}
    
    \begin{tabular}{cccc}
    \hline
    Redshift Range   & Parameter & $\sigma_{\mathrm{stat}}$ & $\sigma_{\mathrm{total}}$ \\ \hline
    \rule{0pt}{2.75ex}\multirow{ 2}{*}{$0.07 < z < 0.2$} & $f \sigma_8$ & $^{+0.16}_{-0.23}$ & $^{+0.16}_{-0.23}$ \vspace{0.1cm}\\ 
    & $\alpha_\perp / \alpha_\parallel$ & $^{+0.044}_{-0.052}$ & $^{+0.045}_{-0.053}$  \vspace{0.2cm}\\
    \multirow{ 2}{*}{$0.2 < z < 0.3$}  & $f \sigma_8$ & $^{+0.14}_{-0.16}$ & $^{+0.14}_{-0.16}$ \vspace{0.1cm}\\
     & $\alpha_\perp / \alpha_\parallel$ & $^{+0.028}_{-0.028}$ & $^{+0.029}_{-0.029}$ \vspace{0.2cm}\\
    \multirow{ 2}{*}{$0.3 < z < 0.4$}  & $f \sigma_8$ & $^{+0.11}_{-0.11}$ & $^{+0.11}_{-0.11}$ \vspace{0.1cm}\\
     & $\alpha_\perp / \alpha_\parallel$  & $^{+0.024}_{-0.024}$ & $^{+0.025}_{-0.025}$ \vspace{0.2cm}\\
    \multirow{ 2}{*}{$0.4 < z < 0.5$}  & $f \sigma_8$ & $^{+0.10}_{-0.10}$ & $^{+0.10}_{-0.10}$ \vspace{0.1cm}\\
    & $\alpha_\perp / \alpha_\parallel$ & $^{+0.020}_{-0.020}$ & $^{+0.021}_{-0.021}$ \vspace{0.2cm}\\
    \multirow{ 2}{*}{$0.5 < z < 0.6$}  & $f \sigma_8$ & $^{+0.084}_{-0.084}$ & $^{+0.086}_{-0.086}$ \vspace{0.1cm}\\
     & $\alpha_\perp / \alpha_\parallel$ & $^{+0.019}_{-0.019}$ & $^{+0.020}_{-0.020}$ \\[1mm] \hline
    \end{tabular} 
    \end{center}
\end{table}

\section{Results}\label{sec:result}

\begin{table}
\centering
\caption{Marginalised 1D constraints on $f\sigma_8$ and $D_\mathrm{M}/D_\mathrm{H}$ at different redshifts, and their correlation coefficient $\rho$.
}
\label{tab:results}
\begin{tabular}{ccccc}
\hline
$z_{\mathrm{eff}}$ & $f \sigma_8$  & $D_\mathrm{M}/D_\mathrm{H}$  & $\rho$ \\ \hline
\rule{0pt}{2.5ex}$0.15$             & $0.51^{+0.16}_{-0.23}$ & $0.156^{+0.007}_{-0.008}$   & $-0.351$               \vspace{0.1cm}\\
$0.26$             & $0.44^{+0.14}_{-0.16}$ & $0.273^{+0.008}_{-0.008}$ & $-0.293$ \vspace{0.1cm}\\
$0.35$             & $0.33^{+0.11}_{-0.11}$ & $0.397^{+0.009}_{-0.009}$ & $-0.287$   \vspace{0.1cm}\\
$0.47$             & $0.53^{+0.1}_{-0.1}$ & $0.556^{+0.011}_{-0.011}$ & $-0.158$   \vspace{0.1cm}\\
$0.54$             & $0.64^{+0.077}_{-0.077}$ & $0.642^{+0.012}_{-0.012}$ & $-0.158$\vspace{0.1cm}\\
$0.69$\parnote{From \citet{Nadathur:2020b}} & $0.356^{+0.079}_{-0.079}$ & $0.868^{+0.017}_{-0.017}$ & $-0.154$   \\[1mm] \hline
\end{tabular}
\parnotes
\end{table}

\begin{figure*}
\centering
\includegraphics[width=.41\textwidth]{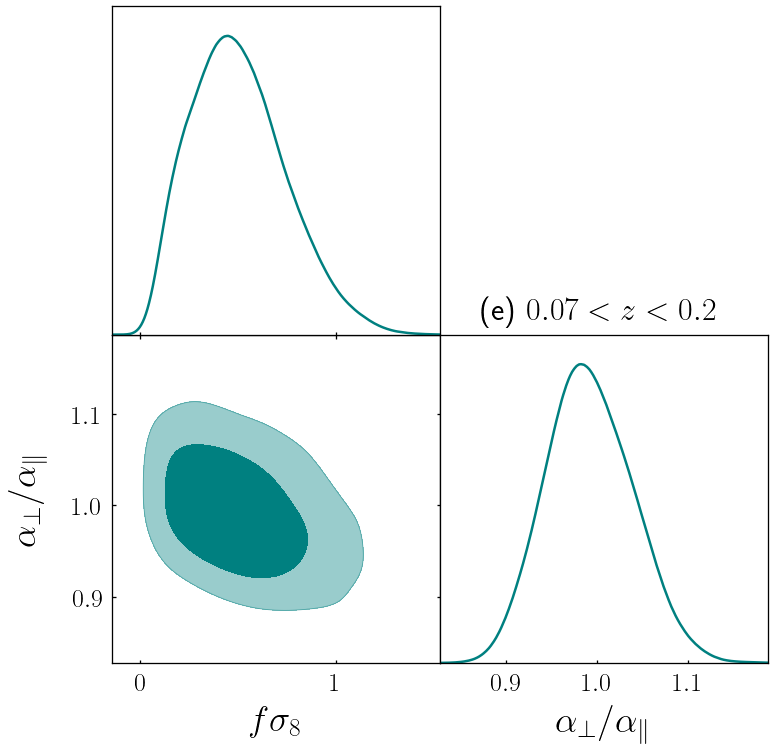}\quad\\
\medskip
\includegraphics[width=.41\textwidth]{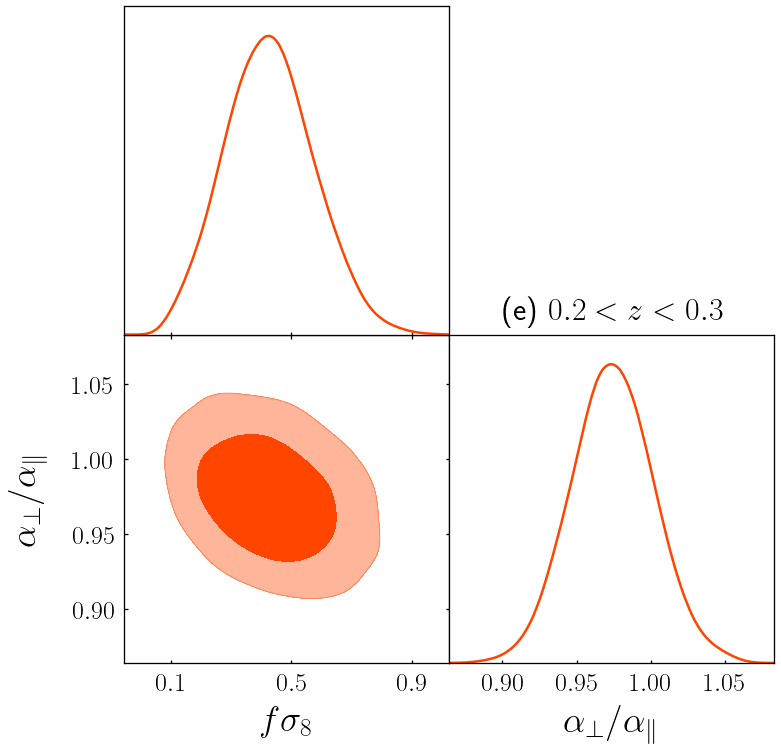}\quad
\includegraphics[width=.41\textwidth]{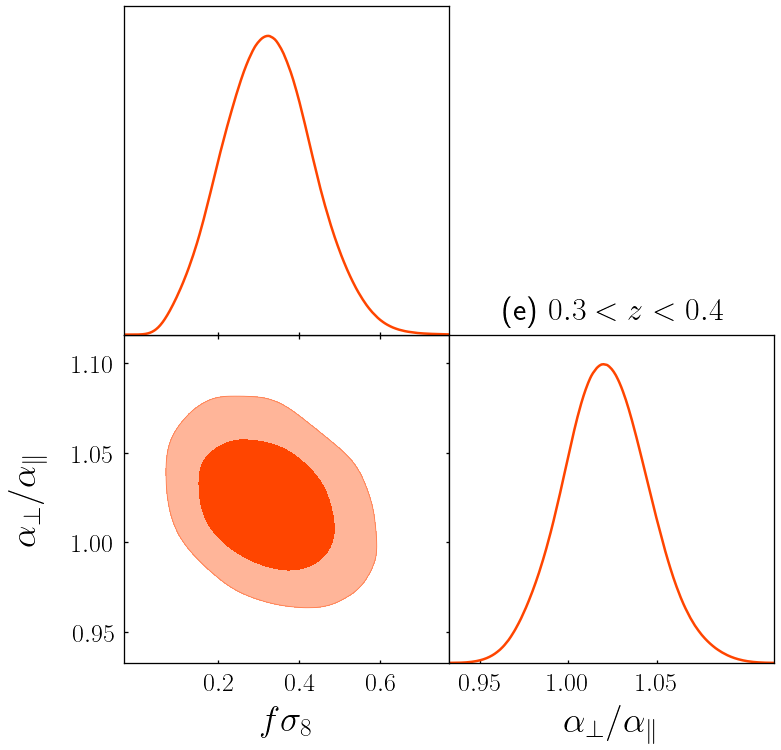}
\medskip
\includegraphics[width=.41\textwidth]{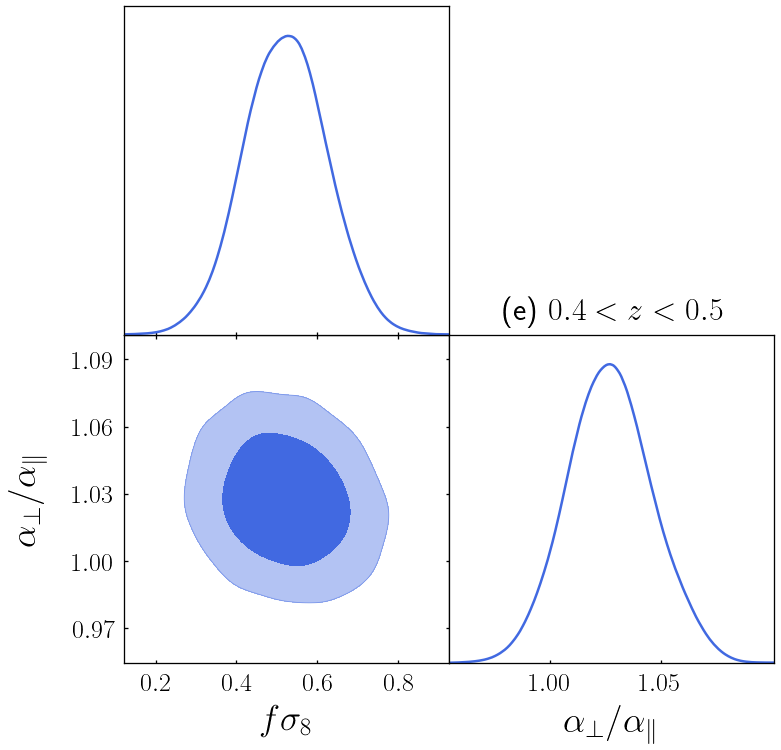}\quad
\includegraphics[width=.41\textwidth]{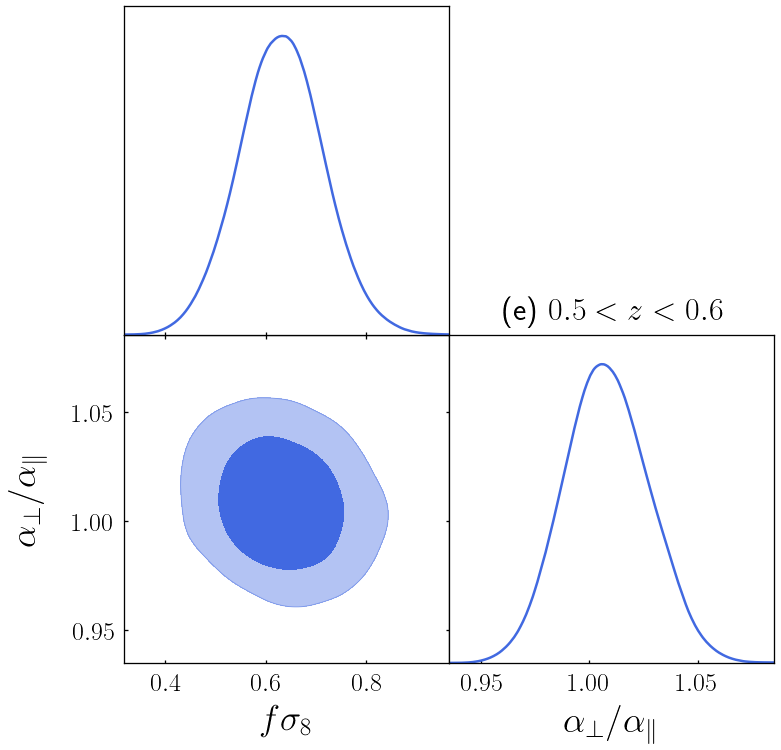}

\caption{Marginalised posterior constraints on the cosmological parameters $f\sigma_8$ and $\alpha_\perp / \alpha_\parallel$ from the fit to MGS and BOSS data in the different redshift bins from Table~\ref{tab:zeff}. Panel (a) (top row) is for MGS, panels (b) and (c) (middle row) are for the two BOSS LOWZ bins, and panels (d) and (e) (bottom row) are for BOSS CMASS, with the colours for the samples matching those in Figure~\ref{fig:footprints}. Shaded contours show the 68\% and 95\% confidence limit regions. These plots include only statistical errors. }
\label{fig:data}
\end{figure*}

\begin{figure}
    \centering
    \includegraphics[width=0.45\textwidth]{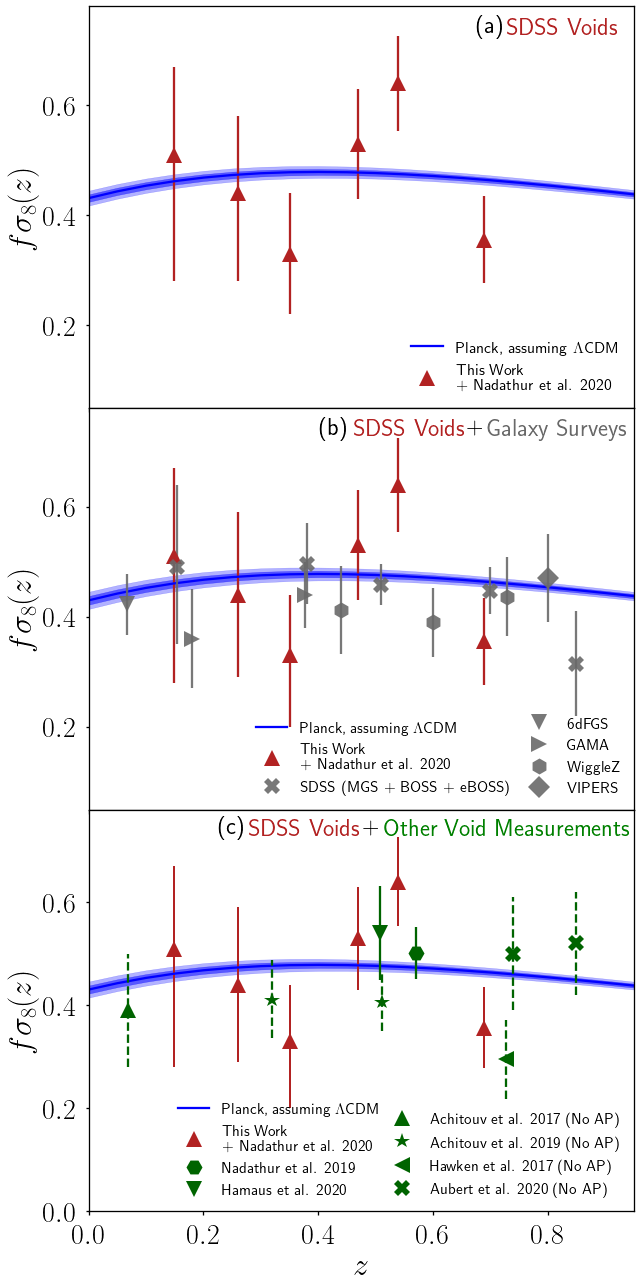}
    \caption{Top: Results for the growth rate as function of redshift, $f(z)\sigma_8(z)$, determined from measurement of the void-galaxy cross-correlation in the MGS and BOSS data presented in this work and in eBOSS from \citet{Nadathur:2020b}. The blue line and shaded regions represent the 68\% and 95\% confidence limits derived from extrapolating CMB measurements from Planck down to these redshift assuming a $\Lambda$CDM model. 
    Centre: As above, but with results obtained from the galaxy clustering and power spectrum from various surveys shown in red. Galaxy survey results are shown from 6dFGS \citep{Beutler2012}, GAMA \citep{Blake2013}, WiggleZ \citep{Blake2012}, VIPERS \citep{delaTorre2013}, MGS \citep{Howlett:2015}, BOSS \citep{Alam-BOSS:2017}, and eBOSS \citep{Alam-eBOSS:2021}. 
    Bottom: As in the top panel, but showing in green results obtained from other analyses of the void-galaxy correlation using alternative analysis techniques, from VIPERS \citep{Hawken:2017}, 6dFGS \citep{Achitouv2017}, multiple re-analyses of BOSS \citep{Nadathur:2019c, Hamaus2020, Achitouv:2019}, eBOSS LRG voids \citep{Aubert20a}, and eBOSS ELG voids \citep{Aubert20a}. Points shown with dashed errorbars are from studies that fix the cosmological model and do not marginalise over the Alcock-Paczynski parameter in reporting growth constraints. }
    \label{fig:fs8-red} 
\end{figure}

Having quantified the contribution of systematic errors through analysis of the mocks, we now turn to the SDSS data. We run our fitting pipeline on the data in each redshift bin exactly as described above for the mocks, using a fiducial cosmology with $\Omega_\mathrm{m}=0.31$. Figure~\ref{fig:bestfit} shows the comparisons between the measured void-galaxy correlation in each bin and the corresponding best-fit model obtained from the fit, together with shading indicating the one standard deviation range of the mock data for the same redshift bin. The resultant marginalized constraints (including only statistical errors) on $f \sigma_8$ and $\alpha_\perp / \alpha_\parallel$ are shown in Figure~\ref{fig:data}. 

The most likely parameter values from the fit to the data, along with the associated statistical error, are also displayed in Figure~\ref{fig:sdss_scatter} for comparison with the scatter seen in the fit to the mocks. We find that the scatter in the mock results is consistent with the mean value and statistical error derived from the MCMC analysis of the data. Table~\ref{tab:total error} summarises the statistical and systematic contribution to the total marginalised 1D errors on $f \sigma_8$ and $\alpha_\perp / \alpha_\parallel$ in each redshift bin.

Our final results are presented in Table~\ref{tab:results}. Here we have converted the measurements of $\alpha_\perp / \alpha_\parallel$ to values for the cosmological distance ratio $D_\mathrm{M} / D_\mathrm{H}$ at each redshift using the values of $D_\mathrm{M}^{\mathrm{fid}}$ and $D_\mathrm{H}^{\mathrm{fid}}$ in the fiducial cosmology. We present the 1D marginalised mean and uncertainty on $f \sigma_8$ and $D_\mathrm{M} / D_\mathrm{H}$ individually, and also the correlation coefficient for their uncertainties, estimated from the statistical errors in the MCMC fit only (as the systematic errors are assumed uncorrelated, Section~\ref{sec:systematic}). It is apparent that the fit values of $f \sigma_8$ and $D_\mathrm{M} / D_\mathrm{H}$ are negatively correlated, with correlation decreasing with increasing redshift.

Figure \ref{fig:fs8-red} displays the measurements of the growth rate $f \sigma_8$ obtained here in comparison to other observational results in the same range of redshifts. Panel (a) (top) compares results from the void-galaxy measurements in this work and \citet{Nadathur:2020b} to those extrapolated to low redshifts from a fit to the Planck CMB data, assuming a flat $\Lambda$CDM cosmology. Panel (b) (middle) compares our results to those measured from standard galaxy clustering techniques without voids obtained from SDSS as well as other surveys in the same range of redshifts. The precision we obtain on $f\sigma_8$ from voids alone is comparable to that from galaxy clustering, and \citet{Nadathur:2019c, Nadathur:2020b} have shown how these two approaches can be consistently combined to obtain more precise measurements than for either alone. 

Finally, in panel (c) of Figure \ref{fig:fs8-red} we compare our growth rate measurements to those obtained from a number of other void-galaxy analyses in the literature. Where these literature results have been reported only in terms of constraints on $\beta=f/b$ \citep{Achitouv:2019} for the purposes of comparison we have translated these values to equivalent constraints on $f\sigma_8$ assuming perfect knowledge of the fiducial galaxy bias $b$ \citep[taking $b=1.85$ for LRGs][]{Alam-BOSS:2017}, and with $\sigma_8(z)$ obtained from extrapolating the central value from \citet{Planck2020}. Several previous void-galaxy analyses \citep[e.g.][]{Achitouv2017, Hawken:2017, Achitouv:2019, Aubert20a} performed fits for the RSD contributions only, with the value of $D_\mathrm{M}/D_\mathrm{H}$ being fixed to that in the fiducial cosmology. Given the correlation between these two parameters (Table~\ref{tab:results}), fixing the cosmology in this way will lead to an underestimate of the marginalised uncertainty in $f\sigma_8$, so the published uncertainties should be treated as lower bounds only. We show the error bars for these studies with dashed lines in the figure in order to highlight this caveat. 

\begin{figure}
    \centering
    \includegraphics[width=0.47\textwidth]{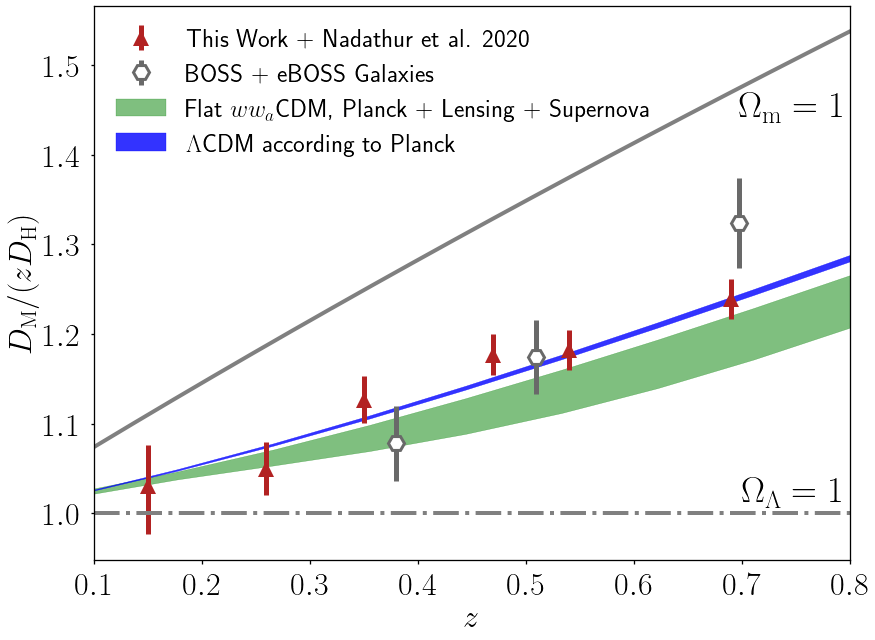}
    \caption{\label{fig:DMDH-red} Measurements of $D_\mathrm{M}/D_\mathrm{H}$, divided by the redshift $z$, from voids in this work are shown as the red triangles with associated error bars. Open grey points show the corresponding results from \citet{Alam-BOSS:2017,Alam-eBOSS:2021} obtained using BAO measured in the same galaxy samples where applicable (transverse and perpendicular BAO were not separately constrained for MGS at $z=0.15$). The blue shaded band is the $68\%$ C.L. region obtained from extrapolating the Planck CMB constraints to low redshifts assuming $\Lambda$CDM. The green shaded band shows the $68\%$ C.L. region from fits to Planck CMB and CMB lensing, and Pantheon supernovae in the $ww_a$CDM model with varying DE equation of state. The grey solid and dot-dashed lines show the expectation for a flat pure matter Universe ($\Omega_\mathrm{m} = 1, \Omega_\Lambda = 0$, $w = -1$) and a flat pure dark energy Universe ($\Omega_\mathrm{m} = 0, \Omega_\Lambda = 1$, $w = -1$) respectively. }
\end{figure}

Figure~\ref{fig:DMDH-red} summarises our results on the background expansion, showing our measurements and marginalised uncertainties for the distance ratio $D_\mathrm{M}/ D_\mathrm{H}$ divided by redshift $z$ for visual clarity, as a function of $z$. Also shown for context are the expectations for two example flat models, with $\Omega_\mathrm{m}=1$, $\Omega_\Lambda=0$ and $\Omega_\mathrm{m}=0$, $\Omega_\Lambda=1$ respectively. For comparison, the grey points we show the equivalent constraints obtained on this quantity from measurement of the BAO signal in the BOSS and eBOSS LRG samples \citep{Alam-eBOSS:2021}. The ratio $D_\mathrm{M}/D_\mathrm{H}$ can only be measured by anisotropic fits to the BAO, which were not possible for the SDSS MGS sample at low redshift \citep{Ross:2015} or the emission line galaxy sample at $z=0.85$ \citep{Raichoor:2021, deMattia:2021}. The blue band indicates the 68\% confidence region obtained from a flat $\Lambda$CDM model fit to the Planck CMB data extrapolated down to low redshifts. The green band shows the same thing for the $ww_a$CDM extended dark energy model but where the fit now includes Planck CMB temperature, polarization and lensing, and Pantheon type Ia supernova data. 
This differs quite markedly from the blue band because of the known slight preference of the Planck data for a dark energy equation of state $w\neq-1$. Figure~\ref{fig:DMDH-red} makes clear the role that the geometrical void-galaxy correlation measurements of $D_\mathrm{M}/D_\mathrm{H}$ at low redshifts can have in distinguishing models of late-time acceleration.

\begin{figure}
    \centering
\includegraphics[width=0.47\textwidth]{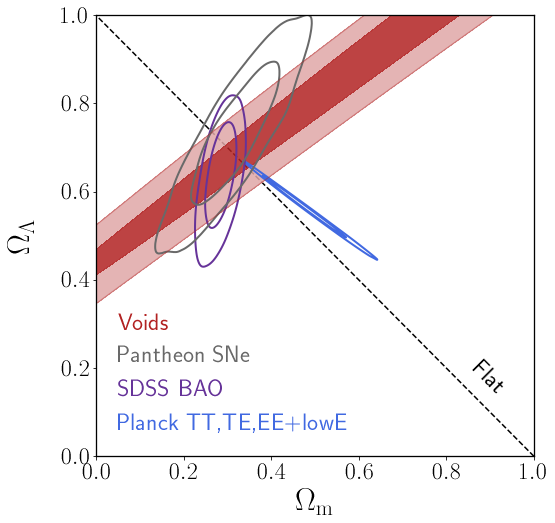}
\caption{\label{fig:OMOL_marginalized} Marginalized constraints on $\Omega_\mathrm{m}$ and $\Omega_\Lambda$, assuming $w=-1$, obtained from void information of this work alone, Pantheon SNe \citep{Scolnic2018}, SDSS BAO \citep{Alam-eBOSS:2021}, and Planck CMB temperature and polarisation \citep{Planck2020}. The black line indicates spatially flat models.}
\end{figure}

As a further illustration of the power of our measurements of $D_\mathrm{M}/D_\mathrm{H}$ for cosmology, in Figure~\ref{fig:OMOL_marginalized} we show the constraints obtained on a non-flat model with free $\Omega_\mathrm{m}$ and $\Omega_\Lambda$ but fixed dark energy equation of state $w=-1$, commonly referred to as the o$\Lambda$CDM model \citep[e.g.][]{Alam-eBOSS:2021}. Figure~\ref{fig:OMOL_marginalized} shows the marginalised posterior constraints for this model in the $\Omega_\mathrm{m}$-$\Omega_\Lambda$ plane obtained from our void-galaxy results for $D_\mathrm{M}/D_\mathrm{H}$ in 6 redshift bins, compared to those obtained from the Planck CMB temperature and polarisation data \citep{Planck2020}, Pantheon SNIa \citep{Scolnic2018} and SDSS BAO measurements from galaxies, quasars and the Lyman-$\alpha$ forest \citep{Alam-eBOSS:2021}. The constraint from voids appears as a narrow band in the $\Omega_\mathrm{m}$-$\Omega_\Lambda$ plane. Assuming only that $\Omega_\mathrm{m}$ cannot be negative, we find that our void measurements alone show very strong evidence for dark energy and accelerated expansion, requiring that $\Omega_\Lambda>0$ at around the $8.7\sigma$ level (determined from the posterior for $\Omega_\Lambda$ at fixed $\Omega_\mathrm{m}=0$). 

A similar band-like degeneracy in the $\Omega_\mathrm{m}$-$\Omega_\Lambda$ plane was obtained by \citet{Nadathur:2020a} when using only one measurement of $D_\mathrm{M}/D_\mathrm{H}$ from voids at $z=0.57$. From Eq.~\ref{eq:DM} it follows that the Alcock-Paczysnki parameter $D_\mathrm{M}/D_\mathrm{H}$ that we measure here depends on the curvature $\Omega_\mathrm{K}=1-\Omega_\mathrm{m}-\Omega_\Lambda$ and the normalized expansion rate $H(z)/H_0$.  For models with a constant dark energy equation of state $w=-1$, this gives rise to a locus of models in the $\Omega_\mathrm{m}$-$\Omega_\Lambda$ plane which have constant $D_\mathrm{M}/D_\mathrm{H}$ at a given redshift. Thus measurement of $D_\mathrm{M}/D_\mathrm{H}$ at a single redshift corresponds to a perfect degeneracy between $\Omega_\mathrm{m}$ and $\Omega_\Lambda$. With $\Omega_\Lambda = a + b\Omega_\mathrm{m}$ for arbitrary $a$ and $b$, requiring a constant value of $D_\mathrm{M}/D_\mathrm{H}$ at redshift $z=0.15$ sets the local gradient $b=0.58$, while for example at redshift $z=0.69$ this translates to $b=0.81$. This indicates that in principle measurements of $D_\mathrm{M}/D_\mathrm{H}$ made at sufficiently many widely separated redshifts can break the $\Omega_\mathrm{m}$-$\Omega_\Lambda$ degeneracy in this class of models. The precision obtained in our current results covering the redshift range $0.15\leq z\leq0.69$ is not sufficient to fully break this degeneracy and close the void-only contours shown, but we expect that the inclusion of more data at higher redshifts from the DESI and Euclid surveys will help to achieve this. Alternatively, since the degeneracy direction is different to that obtained from BAO or SNIa data, one can combine void measurements with these complementary probes to obtain significantly tighter constraints on dark energy, as done by \citet{Nadathur:2020a}.

Further imposing the assumption of flatness ($\Omega_\mathrm{m}+\Omega_\Lambda=1$) allows measurement of $D_\mathrm{M}/D_\mathrm{H}$ to be directly translated to constraints on the single parameter $\Omega_\mathrm{m}$. In this case we find that our results from voids alone result in $\Omega_\mathrm{m}=0.337^{+0.026}_{-0.029}$.

\section{Conclusions}
\label{sec:conclusions}

This work presents a cosmological analysis of the anisotropic void-galaxy cross-correlation measured over a wide redshift range in the SDSS DR7 (MGS) and SDSS DR12 (BOSS LOWZ and BOSS CMASS) galaxy surveys. Voids are extracted after running reconstruction-based RSD removal on the galaxy field in order to remove systematic void selection bias effects. This is implemented in the \texttt{Revolver} code along with \texttt{ZOBOV}, a watershed based void-finder. We performed a joint fit to the multipoles of the measured correlation to determine the growth rate of structure $f\sigma_8$ and the Alcock-Paczynski distance ratio $D_\mathrm{M}/D_\mathrm{H}$ in five redshift bins. Our methods are consistent with those used in previous analyses by \cite{Nadathur:2019c,Nadathur:2020b}, with only minor differences in treatment of uncertainties in the estimated covariance in the likelihood. We therefore combine the measurements performed in this work with the results presented by \citet{Nadathur:2020b} using the eBOSS DR16 data at redshift $z>0.6$ to build a consistent picture of the growth of structure and geometrical evolution of the Universe using voids alone in six independent redshift bins jointly covering the range $0.15\leq z\leq0.69$. Our final results for $f\sigma_8$ and $D_\mathrm{M}/D_\mathrm{H}$ are summarized in Table~\ref{tab:results} and are in excellent agreement with the standard flat $\Lambda$CDM cosmological model.  

We used a large suite of mock galaxy surveys---constructed using both full $N$-body simulations and approximate gravity solvers---to perform multiple tests of possible systematic errors in our analysis in Section~\ref{sec:systematic}. These are quantified as part of the total error budget reported for our results. We found that systematic errors are always small compared to the statistical uncertainties from the data, ensuring that the method and results presented here are robust. Nevertheless, to make full use of the much greater statistical precision that is expected from the much larger datasets that will be available from the DESI and Euclid surveys in the near future, further improvements on the method presented here will be required.

We note that mock catalogues used in this work rely on a HOD to place galaxies in dark matter halos. This HOD is consistent across all environments in the mock and is not adjusted based on whether a galaxy is being placed in a high density region or a low density region (such as a void). \citet{Tinker2006,Tinker2008,Tinker2009} show that galaxy-halo connection shows no strong changes in low-density environments such as voids. In contrast, \citet{Verza2022} find a scale dependence for halo bias inside voids. Effects of other prescriptions for HODs will be tested in future work. 

Our work shows the importance of voids as cosmological probes and motivates the inclusion of voids as standard tools in the analysis of galaxy survey data due to the information gain available from void-galaxy correlations. This is particularly relevant to low-redshift geometrical tests of cosmological expansion and discriminating between alternative models of dark energy that change the expansion history at late times. While our current results are in excellent agreement with the flat $\Lambda$CDM model, in the near future DESI and Euclid will probe much larger volumes of the Universe over a larger redshift range and constraints from voids in these surveys will provide a powerful test of cosmological models building on the results in this work.

\section*{Acknowledgments}

AW acknowledges the support of the Natural Sciences and Engineering Research Council of Canada (NSERC) [funding reference number 547744]. Cette recherche a été financée par le Conseil de recherches en sciences naturelles et en génie du Canada (CRSNG) [numéro de référence 547744]. SN acknowledges support from an STFC Ernest Rutherford Fellowship, grant reference ST/T005009/2.

Research at Perimeter Institute is supported in part by the Government of Canada through the Department of Innovation, Science and Economic Development Canada and by the Province of Ontario through the Ministry of Colleges and Universities.

This research was enabled in part by support provided by Compute Ontario (computeontario.ca) and Compute Canada (computecanada.ca).

Funding for the Sloan Digital Sky Survey IV has been provided by the Alfred P. Sloan Foundation, the U.S. Department of Energy Office of Science, and the Participating Institutions. 

SDSS-IV acknowledges support and resources from the Center for High Performance Computing  at the University of Utah. The SDSS website is www.sdss.org.

SDSS-IV is managed by the Astrophysical Research Consortium for the Participating Institutions of the SDSS Collaboration including the Brazilian Participation Group, the Carnegie Institution for Science, 
Carnegie Mellon University, Center for Astrophysics | Harvard \& Smithsonian, the Chilean Participation 
Group, the French Participation Group, Instituto de Astrof\'isica de Canarias, The Johns Hopkins University, Kavli Institute for the Physics and Mathematics of the Universe (IPMU) / University of Tokyo, the Korean Participation Group, Lawrence Berkeley National Laboratory, Leibniz Institut f\"ur Astrophysik Potsdam (AIP),  Max-Planck-Institut f\"ur Astronomie (MPIA Heidelberg), Max-Planck-Institut f\"ur Astrophysik (MPA Garching), Max-Planck-Institut f\"ur Extraterrestrische Physik (MPE), National Astronomical Observatories of China, New Mexico State University, New York University, University of Notre Dame, Observat\'ario Nacional / MCTI, The Ohio State University, Pennsylvania State University, Shanghai Astronomical Observatory, United Kingdom Participation Group, Universidad Nacional Aut\'onoma 
de M\'exico, University of Arizona, University of Colorado Boulder, University of Oxford, University of Portsmouth, University of Utah, University of Virginia, University of Washington, University of Wisconsin, Vanderbilt University, and Yale University.

For the purpose of open access, the authors have applied a CC BY public copyright licence to any Author Accepted Manuscript version arising.

\section*{Data Availability Statement}

Data supporting this research including the measured correlation functions, covariance matrices, and resulting likelihood for cosmological parameters are available on request from the corresponding author and will be made public in the \texttt{Victor} repository at \url{https://github.com/seshnadathur/victor} upon publication.



\bibliographystyle{mnras}
\bibliography{paper}






\bsp	
\label{lastpage}
\end{document}